\documentclass[twocolumn]{IEEEtran} 
\usepackage[T1]{fontenc}
\usepackage{color}
\usepackage{float}
\usepackage{mathrsfs}
\usepackage{amsmath}
\usepackage{amsthm} 
\usepackage{amssymb}
\usepackage{graphicx}
\usepackage{cite}
\usepackage{makecell}
\usepackage{fancyhdr}
\usepackage{diagbox}
\usepackage{bm}
\usepackage{CJKutf8}
\usepackage[ruled,linesnumbered]{algorithm2e} 
\usepackage[unicode=true,
bookmarks=true,bookmarksnumbered=true,bookmarksopen=true,bookmarksopenlevel=1,
breaklinks=false,pdfborder={0 0 0},pdfborderstyle={},backref=false,colorlinks=false]
{hyperref}
\hypersetup{
pdftitle={Your Title},
pdfauthor={Your Name},
pdfpagelayout=OneColumn, pdfnewwindow=true, pdfstartview=XYZ, plainpages=false}

\usepackage[caption=false,font=footnotesize]{subfig} 
\usepackage{multirow} 
\usepackage{xcolor}

\allowdisplaybreaks[4]

\ifCLASSOPTIONcompsoc
 \usepackage[caption=false,font=normalsize,labelfont=sf,textfont=sf]{subfig}
\else
 \usepackage[caption=false,font=footnotesize]{subfig}
\fi

\IEEEoverridecommandlockouts

\usepackage{geometry}
\begin{document}
\begin{CJK}{UTF8}{gbsn}
\title{Three-Dimension Collision-Free Trajectory Planning of UAVs Based
on ADS-B Information in Low-Altitude Urban Airspace}
\author{Chao Dong, Yifan Zhang, Ziye Jia, Yiyang Liao, Lei Zhang, and Qihui Wu
\thanks{
This work was supported in part by the National Key R\&D Program of China 2022YFB3104502, in part by National Natural Science Foundation of China under Grant 62301251, in part by the Natural Science Foundation of Jiangsu Province of China under Project BK20220883, in part by the open research fund of National Mobile Communications Research Laboratory, Southeast University (No. 2024D04), and  in part by the Young Elite Scientists Sponsorship Program by CAST 2023QNRC001.
\par Chao Dong, Yifan Zhang, Yiyang Liao, Lei Zhang and Qihui Wu are with the College of Electronic and Information Engineering, Nanjing University of Aeronautics and Astronautics,
Nanjing 211106, China (e-mail: dch@nuaa.edu.cn; yifanzhang123@nuaa.edu.cn; liaoyiyang@nuaa.edu.cn;  Zhang\_lei@nuaa.edu.cn; wuqihui@nuaa.edu.cn;).
\par Ziye Jia is with the College of Electronic and Information Engineering, Nanjing University of Aeronautics and Astronautics, Nanjing 211106, China, and also with the National Mobile Communications Research Laboratory, Southeast University, Nanjing 211111, China (e-mail: jiaziye@nuaa.edu.cn).
\par \emph{Corresponding author: Ziye Jia. }}}
\maketitle

\begin{abstract}
The environment of low-altitude urban airspace is complex and variable
due to numerous obstacles, non-cooperative aircrafts, and birds. Unmanned
aerial vehicles (UAVs) leveraging environmental information to achieve
three-dimension collision-free trajectory planning is the prerequisite
to ensure airspace security. However, the timely information of surrounding
situation is difficult to acquire by UAVs, which further brings security
risks. As a mature technology leveraged in traditional civil aviation,
the automatic dependent surveillance-broadcast (ADS-B) realizes continuous surveillance of the information of aircrafts. Consequently,
we leverage ADS-B for surveillance and information broadcasting, and
divide the aerial airspace into multiple sub-airspaces to improve
flight safety in UAV trajectory planning. In detail, we propose the
secure sub-airspaces planning (SSP) algorithm and particle swarm optimization
rapidly-exploring random trees (PSO-RRT) algorithm for the UAV trajectory
planning in law-altitude airspace. The performance of the proposed
algorithm is verified by simulations and the results show that SSP
reduces both the maximum number of UAVs in the sub-airspace and the
length of the trajectory, and PSO-RRT reduces the cost of UAV trajectory
in the sub-airspace.
\end{abstract}


\section{Introduction}

\IEEEPARstart{W}{ith} the advantages of high mobility and low cost,
unmanned aerial vehicles (UAVs) are capable of many tasks such as
air surveillance \cite{zhu2023uav,9228876,9377560}, freight delivering
\cite{10197535}, auxiliary communication\cite{9184929,9718359} and
computation\cite{9714482}, and disaster rescue \cite{9166731}. Compared
with other ground vehicles, the trajectories of UAVs are more flexible
\cite{9679716}. Besides, UAVs are able to select efficient trajectories
to complete required tasks. UAVs are extensively utilized in low-altitude
urban airspace owing to their cost-effectiveness, adaptability, and
maneuverability. However, due to the limitation of endurance of UAVs,
it is necessary to plan a collision-free trajectory within the energy
constraints \cite{9258936}. Furthermore, the incorrect acquisition
of position information of UAVs may cause collisions with obstacles,
which is unacceptable for low-altitude urban airspaces. The low-altitude
urban airspace is characterized by a complex and variable environment
\cite{8766890}, featuring unforeseen events such as birds and non-cooperative
UAVs. Due to limited environment perception of UAVs, the ground-assisted-airspace
safety assessment becomes imperative, which requires swift information
exchange among UAVs and ground surveillance agencies. Considering
the above factors, the UAV must strategically plan a safe and viable
trajectory within the energy constraints to fulfill the assigned task
based on real-time airspace situational information during task execution. 

A well-designed airspace division enhances the efficiency of UAV management
systems. Drawing inspirations from the airspace corridors utilized
by conventional civil aviation aircraft \cite{shanmugavel20073d,ehrmanntraut2007airspace},
the airspace of UAV can also be divided into multiple designated tubes.
By adhering to the pre-planned tubes, UAVs significantly reduce the
probability of collision with obstacles. However, tubes lack flexibility
and struggle to accommodate a large number of UAVs.
An alternative approach to airspace division is stratification, which
vertically segregates the airspace into distinct layers. This approach
allows the airspace to accommodate more UAVs \cite{hoekstra2016layered},
but may compromise safety and operational efficiency \cite{sedov2018centralized}.
Dividing the airspace into discrete grids is another commonly used
method \cite{pang2020concept}. The grid approach allows for sequential
traversal from initial to destination airspace, making it suitable
for UAVs capable of vertical takeoff and landing. Compared with the
tube method, the grid method leads to a higher collision probability,
but this method effectively increases the number of UAVs that can
be accommodated in the airspace. Additionally, it offers greater adaptability
in UAV trajectory selection by refining airspace at the same altitude.
In this paper, in order to make full use of the agility of UAVs, we
divide the low-altitude urban airspace into multiple sub-airspaces.

As a mature technology employed in civil aviation surveillance, the
automatic dependent surveillance broadcast (ADS-B) has the advantages
of fast message update, low cost and rich information, the UAV enhances
its airspace perception\cite{10233390}, enabling the acquisition
of vital information and facilitating applications such as obstacle
avoidance. In this paper, ADS-B device is equipped
for UAVs to enhance the information acquisition and environmental
perception capabilities.

A key prerequisite for UAVs to complete service tasks in low-altitude
urban airspace is to plan a safe trajectory from the starting point
to the endpoint without collisions \cite{9718364}. Traditional trajectory planning methods include the artificial potential
field (APF) \cite{9453807}, A$^\ast$ \cite{cai2019path} and Dijkstra
\cite{ibrahim2009way}, and these methods are widely used in the
trajectory planning of UAV in urban airspace. APF is commonly used
for aircraft trajectory planning. In \cite{qian2023cerebellar},
the area around the destination is set as the gravitational field,
and various types of obstacles are set as the repulsive field to incite
collision during UAVs flight. Both Dijkstra and A$^\ast$ are efficient
in searching the trajectory between the start and the destination
\cite{soltani2002path}. Intelligent algorithm is another way to find
the trajectory in the airspace, such as genetic algorithm (GA), ant
colony optimization (ACO) \cite{li2023uav} and particle swarm optimization
(PSO) \cite{liu2021joint}. GA simulates the genetic mechanism and
natural evolution of organisms in nature. ACO and PSO simulate the
process of ant colony and bird flock to obtain food, respectively. These algorithms use the bionic mechanism of biological
individuals or clusters to find trajectories and avoid collisions,
which are simple to be implemented and have better optimization effect.
As a sampling-based trajectory planning method, rapidly-exploring
random trees (RRT), bidirectional rapidly-exploring random trees (Bi-RRT)
and RRT$^\ast$ are often used in trajectory planning \cite{jiang2021r2,tan2019uav}.
These algorithms find a collision-free trajectory by randomly generating
trajectory points and performing the shortest trajectory update and
timely collision detection. However, much of the
existing research focuses on the two-dimensional trajectory planning
of UAVs, which makes it difficult to fully leverage the high maneuverability
advantage of UAVs. This paper considers the variation of the UAV in
the vertical direction during trajectory planning, making it more
closely aligned with real-world scenarios.

In this work, RRT and Bi-RRT are leveraged as the basic trajectory
search algorithms, which are combined with the PSO algorithm to optimize
the trajectory. The main contributions of this paper are summarized
as follows:
\begin{enumerate}
\item We divide the airspace into grids and utilize ADS-B as information
source for UAVs to obtain airspace status information. Meanwhile,
ground stations broadcast information about sudden obstacles to UAVs
via ADS-B, allowing for trajectory readjustment. 
\item We propose a secure sub-airspaces planning (SSP) algorithm based
on dynamic programming, sliding window, and attraction mechanism for
trajectory planning among sub-airspaces for UAVs. The coarse-grained
trajectory is dynamically adjusted based on the status of airspace,
reducing the maximum number of UAVs in the sub-airspace. 
\item We design the particle swarm optimization-rapidly
random trees (PSO-RRT) algorithm for trajectory planning within the
sub-airspace, which considers both efficiency and cost to ensure safety
and reduce energy consumption in UAV trajectories. The performance
of PSO-RRT is sufficiently demonstrated via simulation results.
\end{enumerate}
\par The organization of this paper is as follows: Section \ref{sec:2}
introduces the related research works. In Section \ref{sec:3}, the
problem expatiation and designed algorithms are presented. Section
\ref{sec:4} provides the simulation results. Finally, the conclusion
is drawn in Section \ref{sec:5}.

\section{Related Works\label{sec:2}}

There exist a couple of researches focusing on UAV trajectory planning
conducted by researchers. In this work, we primarily focus on three
interconnected research fields: the integration of UAVs and ADS-B
systems, airspace design for UAVs and UAV trajectory planning.

ADS-B enhances the situational awareness ability of UAVs in low-altitude
airspace and the surveillance ability of ground stations. \cite{7829019}
studies the cooperative perception and avoidance among UAVs equipped
with ADS-B, proposes a planning algorithm based on RRT, and the simulation
results show that in frontal encounter conflict, the UAV equipped
with the RRT based algorithm successfully realizes the conflict resolution
by leveraging ADS-B. In the context of UAV trajectory prediction,
\cite{10000919} proposes a centralized UAV trajectory surveillance
architecture with ADS-B in low-altitude airspace, and predicts the
ADS-B trajectory. The long short-term memory (LSTM) is leveraged to
train the UAV ADS-B information, and the simulation results reveal
that the proposed algorithm has higher prediction accuracy by leveraging
ADS-B information. In \cite{9077382}, the safety of utilizing ADS-B
in UAVs is investigated, and an algorithm for distinguishing fake
UAV ADS-B information is proposed, ensuring the data security of UAVs.
The utilization of ADS-B for trajectory monitoring and planning on
UAVs shows potential, \textcolor{black}{but there } exist \textcolor{black}{{}
limited researches specifically focusing on utilizing UAV ADS-B data
for trajectory planning.}

In the domain of civil aviation, the division of airspace holds the
potential to augment both flight safety and airspace
utilization for aircrafts. Hence, the airspace division can also enhance
the efficiency and safety of UAVs. \cite{9945667} proposes an airspace
grid division model based on GPS signals and wind strength, which
effectively enhances the utilization of airspace. \cite{9256627}
divides the urban airspace into multiple grids and adjusts their sizes
based on the degree of danger, enabling risk avoidance in UAV trajectory
planning. \cite{he2022route} subdivides the urban airspace into a
series of grids and utilizes a designed cost function for UAV trajectory
planning among grids. The results demonstrate that in various application
scenarios, grids significantly enhance the airspace utilization of
UAVs. \cite{mohamed2018preliminary} models the urban infrastructure
in the three-dimensional airspace and conducts comparative experiments
on trajectory planning using three methods: grid, tube, and trajectory
points. The results demonstrated that compared with
the other algorithms, the grid method has the highest UAV capacity
and throughput. In conclusion, the grid method strikes a balance between
efficiency and safety in UAV trajectory planning.

In the scenario of cargo transportation by UAVs in urban airspace,
\cite{li2021research} redesigns the cost estimation function of A$^\ast$
to enable the planned trajectory to consider both the efficiency and
cost of goods delivery, thus achieving rapid trajectory planning.
In \cite{zhou2022uav}, according to the changes in the airspace,
trajectory planning is performed using the A$^\ast$ and RRT$^\ast$ algorithms,
respectively. When the airspace situation is stable, the UAV utilizes
the A$^\ast$ algorithm. However, when the airspace changes and the original
trajectory becomes invalid, the trajectory is optimized by adjusting
the selection probability and range of trajectory points in the RRT$^\ast$
algorithm to adapt to the changing airspace. Although traditional
methods are easy to implement, the planned trajectories tend to be
rigid, making it difficult to fully leverage the advantages of agile
flight for UAVs.

RRT and its variants efficiently compute collision-free
trajectories within specified airspace. In the context of unknown
environmental information and unavailable GPS signals, \cite{rhodes2022autonomous}
explores the application of RRT$^\ast$ for small UAVs in locating the
source of hazardous chemical leaks. By leveraging the utilization
and exploration mechanism, RRT$^\ast$ generates candidate trajectories
limited to the sensor's sensing range which optimizes computational
resources and enables real-time trajectory planning. \cite{chang2022skeleton}
aims to swiftly determine shorter UAV flight trajectories within the
airspace, this study utilizes the RRT algorithm based on a greedy
approach for trajectory planning to minimize unnecessary bends. The
algorithm reduces search complexity and requires only a few trajectory
points. In densely populated low-altitude airspace with static and
dynamic obstacles, \cite{wen2015uav} models static threats and predicts
dynamic threats using the RRT algorithm. By employing this model,
the RRT$^\ast$ algorithm is utilized for trajectory planning in complex
airspace. The algorithm exhibits a high obstacle penetration rate.
However, the RRT algorithm and variants are known for randomness,
making it challenging to find the optimal trajectory within the airspace
for UAVs.

As an important part of intelligent algorithm, PSO has been applied
in UAV trajectory planning. \cite{yu2022novel} combines simulated
annealing and PSO to realize autonomous trajectory planning of UAVs.
The random disturbance mechanism of simulated annealing algorithm
is used to assist PSO to jump out of local minimum value and avoid
falling into local optimum. The simulation results show that the algorithm
has higher trajectory quality. \cite{roberge2012comparison} leverages
PSO to generate the UAV trajectory in complex three-dimensional environment.
The results show that the PSO algorithm satisfies the requirements
of real-time trajectory planning for UAVs. Based on the above analysis,
PSO demonstrates excellent performance in solving optimization problems
related to UAV trajectories.

\section{Problem Expatiation and Algorithm Design \label{sec:3}}

In this section, the low-altitude urban airspace is divided into multiple
sub-airspaces, and the SSP algorithm is proposed to achieve trajectory
planning among sub-airspaces. The PSO-RRT algorithm is also designed
to achieve trajectory planning within each sub-airspace.

\subsection{Airspace division}

\begin{figure}
\centering

\includegraphics[width=3.5in]{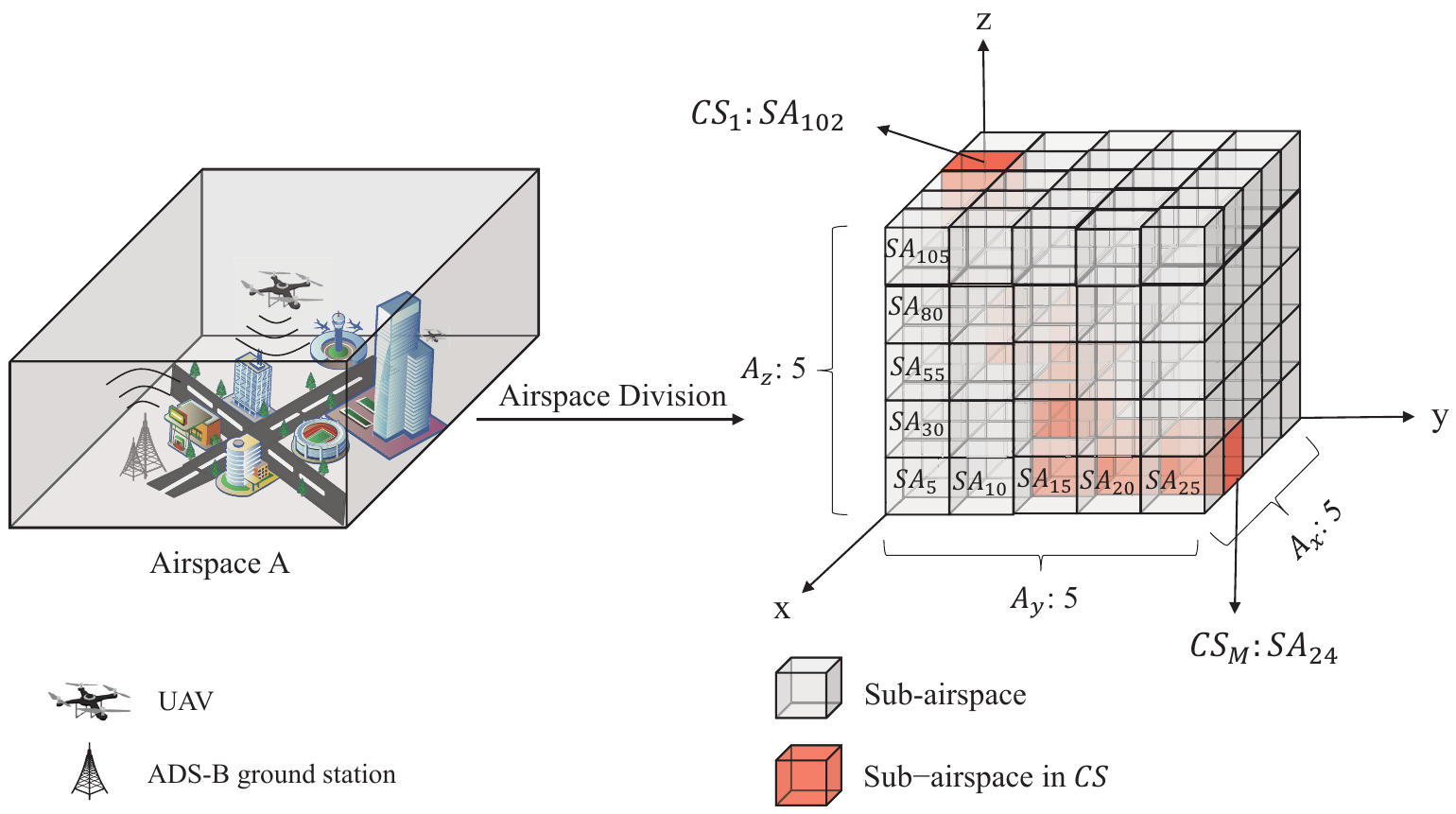}

\vspace{-3mm}
\begin{raggedright}
\caption{\label{fig:1}Scenario and airspace division.}
\par\end{raggedright}
\vspace{-5mm}
\end{figure}

Airspace division is beneficial for UAVs in low-altitude urban airspace.
It enables better warning and avoidance of conflicts with obstacles.
Additionally, when planning UAV trajectories, only obstacles within
the current sub-airspace need to be considered, which reduces complexity
compared to considering all obstacles in the entire airspace. On the
left side of Fig. \ref{fig:1}, the airspace where the UAV $U_{i}$
works is designated as a large area $A$, which contains buildings
of different heights and ADS-B ground stations. UAV $U_{i}$ only
provides services to users within airspace $A$, which means $U_{i}$
will not fly out of the boundaries of $A$. On the right side of Fig.
\ref{fig:1}, $A$ is divided into interconnected, independent, and
equally sized multi-layer grid sub-airspace $\boldsymbol{SA}={SA_{1},\ SA_{2},\ ...,SA_{n},\ ...,\ SA_{N}}$.
$A_{x}$ is the number of sub-airspaces in the $x$ direction, $A_{y}$
is the number of sub-airspaces in the y direction, and $A_{z}$ is
the number of sub-airspaces in the $z$ direction. The number of the
grids is in the order of $x$ direction first and then $y$ direction
layer by layer. In Fig. \ref{fig:1}, the number of sub-airspaces
in all three directions is 5, which means the airspace $A$ is divided
into 125 sub-airspaces.

\subsection{Trajectory planning}

The UAV trajectory planning consists of two main parts. Firstly, the
UAV utilizes SSP for trajectory planning among sub-airspaces to find
continuous coarse-grained trajectory $\boldsymbol{CS}=\{CS_{1},\ CS_{2},\ ...,CS_{m},\ ...,\ CS_{M}\}$,
a coarse-grained trajectory consists of $M$ continuous sub-airspaces.
$CS_{1}$ and $CS_{M}$ are respectively indicate
the starting sub-airspace and the destination sub-airspace. Subsequently,
a fine-grained trajectory planning is performed by PSO-RRT within
each specific sub-airspace in $CS$. This paper assumes that the UAV
possesses positional information of all buildings in $A$ and can
determine its current position coordinates $P_{i}$ using an onboard
positioning system, and the destination coordinates, $P_{e}$. UAVs
in airspace $A$ are equipped with ADS-B IN and ADS-B OUT devices,
allowing for broadcasting and receiving ADS-B messages, respectively.
Since ADS-B can periodically and automatically broadcast
the current positioning information of UAVs, UAVs leverage the position
information broadcasted by other UAVs within the airspace and sudden
obstacle information broadcasted by ground surveillance agencies to
conduct trajectory planning between sub-airspaces and trajectory re-planning
in sub-airspace.

\subsubsection{Trajectory planning among sub-airspaces}

The UAV $U_{i}$ utilizes the SSP algorithm to search for $CS$ and
executes two steps. Firstly, $U_{i}$ determines the starting sub-airspace
$CS_{1}$ and the destination sub-airspace $CS_{M}$ based on coordinating
$P_{i}$ and $P_{e}$. Then, leveraging dynamic programming, $U_{i}$
plans a coarse-grained trajectory $CS$ composed of sub-airspace between
$CS_{1}$ and $CS_{M}$. In Fig. \ref{fig:1}, the $CS_{1}$ and $CS_{11}$
of $U_{i}$ are respectively $SA_{102}$ and $SA_{24}$, and the continuous
red sub-airspaces represent $CS$. $CS$ can achieve a multitude of
combination possibilities by utilizing different sub-airspaces. Therefore,
it is necessary to establish evaluation criterias for comparisons.
In formula (\ref{eq:1}), $CT_{n}$ denotes the cost of $U_{i}$ in
sub-airspace $SA_{n}$, $O_{n}$ denotes the number of static obstacles
in $SA_{n}$, and $AEC_{n}$ denotes the number of UAVs in $SA_{n}$.
$k1$ and $k2$ are polynomials, in particular, $k1+k2=1$, $k1,\;k2>0$,
and $k1\gg k2$:

\begin{equation}
CT_{n}=k1\cdotp O{}_{n}+k2\cdotp AEC_{n}.\label{eq:1}
\end{equation}
The function expressed by formula (\ref{eq:2}) serves as a quantitative
measure for evaluating the performance of different $CS$ configurations:

\begin{equation}
CT_{sn}=\sum_{i=1}^{N}CT_{i}.\label{eq:2}
\end{equation}

Subsequently, the sub-airspaces in $CS$ are optimized in problem
$\mathscr{P}0$ to minimize the associated cost:

\begin{align}
\mathscr{P}0:\;\underset{\boldsymbol{CS}}{\textrm{min\ }} & CT_{sn}\label{eq:3}
\end{align}

\begin{figure}
\centering

\includegraphics[width=3.5in]{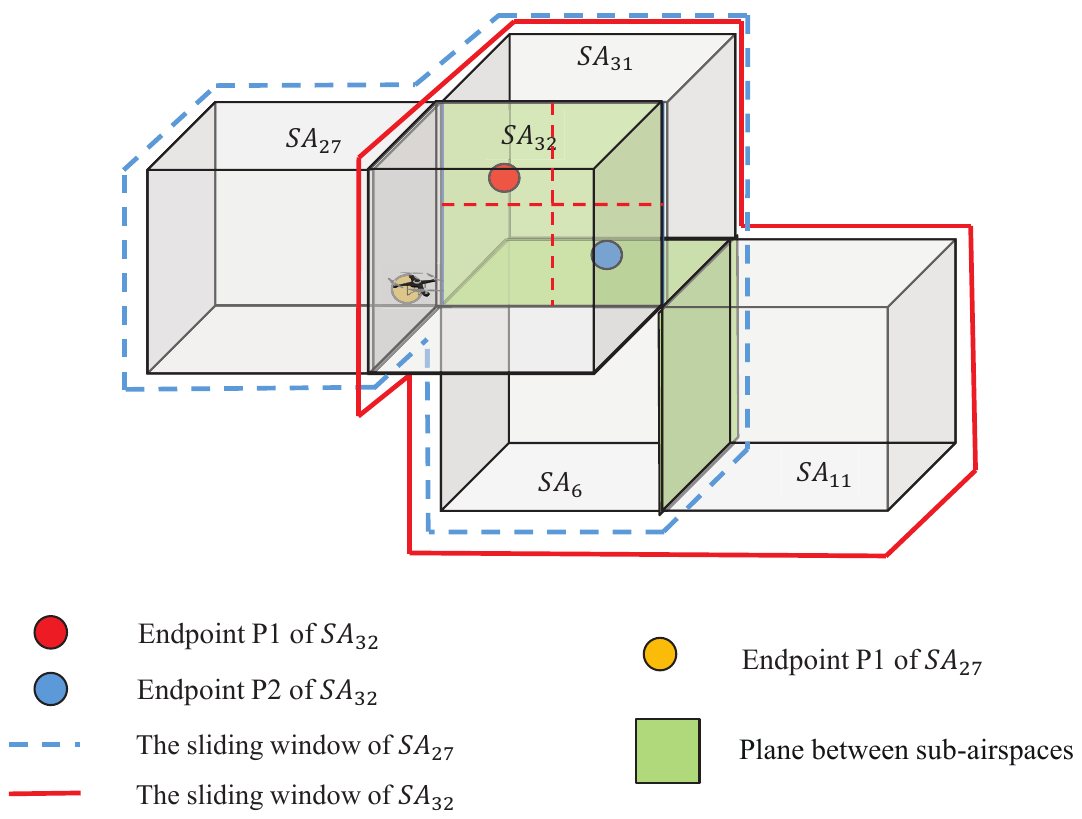}

\vspace{-1mm}
\begin{raggedright}
\caption{\label{fig:2}The sliding window in SSP.}
\par\end{raggedright}
\vspace{-5mm}
\end{figure}

To reduce the probability of collision, a sliding window is designed
as a component in the SSP algorithm. In detail, the working process
of the sliding window for trajectory planning among sub-airspaces
is illustrated in Fig. \ref{fig:2}. When $U_{i}$ enters $SA_{27}$,
formula (\ref{eq:2}) and dynamic programming algorithm are utilized
to obtain the new $CS$. The sub-airspaces $SA_{27}$, $SA_{32}$,
$SA_{31}$, and $SA_{6}$ within the blue dashed area are selected
as the sliding window when $U_{i}$ enters $SA_{32}$. When $U_{i}$
moves from $SA_{27}$ to $SA_{32}$, maintaining $SA_{32}$, $SA_{31}$,
and $SA_{6}$ in the sliding window. formula (\ref{eq:2}) and dynamic
programming algorithm are employed again to obtain the new $CS$.
The sub-airspaces within the red solid box are selected as the sliding
window when UAV $U_{i}$ enters $SA_{32}$. This process of sliding
the sub-airspace window is repeated when UAV $U_{i}$ enters a new
sub-airspace, until there are less than four sub-airspaces remained.
The length for the sliding window needs to be carefully determined.
When the length is too large, it will result in a delay in perceiving
the overall spatial situation, which is not conducive to avoiding
sub-airspaces with a large number of UAVs. Conversely, when the length
is too small, in Fig. \ref{fig:2}, the length of sliding window is
1. When UAV $U_{i}$ enters $SA_{32}$, the endpoint will lie on the
plane between $SA_{32}$ and $SA_{31}$. the position could be $P1$
or $P2$. If the endpoint of $U_{i}$ is $P1$, it leads to a long
path in $SA_{31}$, which means more energy consumption and higher
probability of conflicts. However, when the length of sliding window
is 4, the endpoint $P2$ is determined by attraction mechanism. $P2$
shortens the trajectory in $SA_{31}$, which results
in energy savings and lower probability of conflicts. Therefore, a
window length of 4 is set for SSP to achieve a balance among computational
power, energy consumption, and safety. 

\begin{figure}
\centering

\includegraphics[width=2.8in]{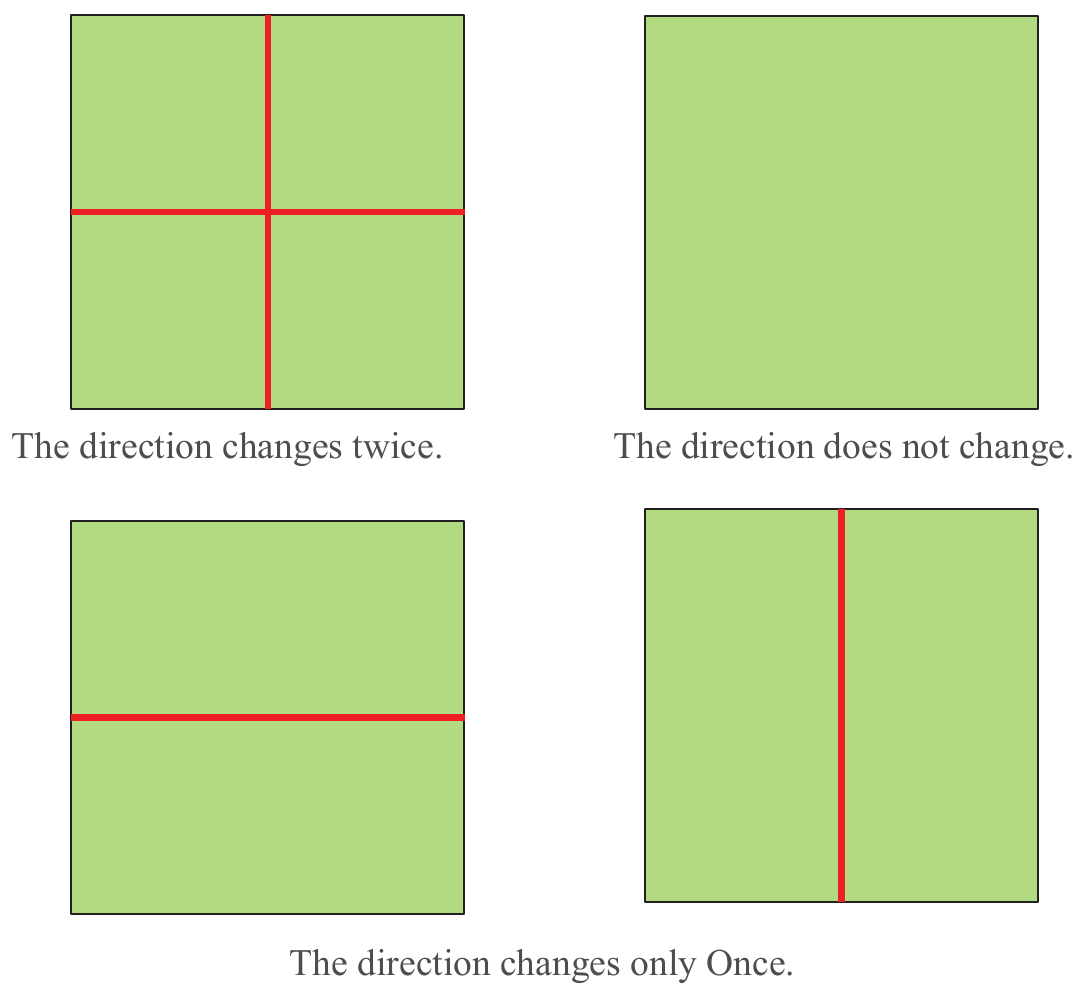}

\vspace{-3mm}

\caption{\label{fig:3}The attraction mechanism in SSP. }

\vspace{-5mm}

\end{figure}

The endpoint position of a sub-airspace will affect the trajectory
length of $U_{i}$. In order to shorten the trajectory length and
save energy consumption, the SSP leverages an attraction mechanism
which attracts the endpoint on the basis of sliding
windows. The plane where the endpoint belongs is
divided into different regions according to the direction of the sub-airspaces
in sliding window. If two directions of the sub-airspace in the sliding
window change, as shown in Fig. \ref{fig:3}, the plane is divided
into four areas. According to the change in the direction of the sub-airspaces,
the endpoint can be limited to one area which shorten the length of
trajectory in next sub-airspace. If only one direction of the sub-airspace
in the sliding window changes, the plane will be divided into two
parts, which still have certain performance improvement compared with
completely randomly finding the endpoint. If the direction has no
change, the selection of the endpoint is completely random, and the
attraction mechanism will lose effect.

\subsubsection{Trajectory planning in sub-airspace}

After finding coarse grained trajectory $CS$ for UAV $U_{i}$, more
precise trajectory planning needs to be performed within the sub-airspace.
In order to make full use of the advantages of PSO
and Bi-RRT, PSO-RRT algorithm is formulated for $U_{i}$ to achieve
quickly locating the trajectory points of UAVs while keeping them
away from obstacles.

\begin{figure}
\centering

\includegraphics[width=3.5in]{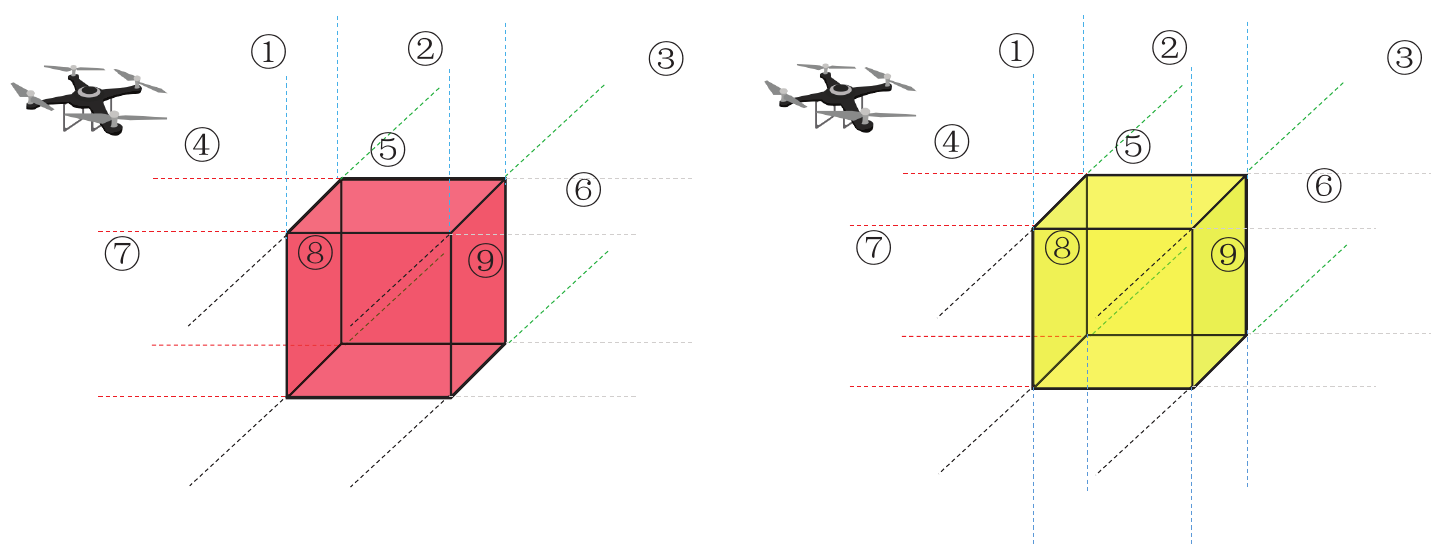}

\vspace{-3mm}

\caption{\label{fig:4}Obstacle models in sub-airspace.}

\vspace{-5mm}
\end{figure}

The distance between UAV $U_{i}$ and obstacles needs to keep a safe
range to ensure safe flight. Therefore, it is necessary to model the
obstacles to calculate the distance between them. The obstacles can
be divided into the static obstacle and sudden obstacle. The red cuboid
in Fig. \ref{fig:4} represents a static obstacle, which mainly includes
ground buildings. The space of static obstacle can be divided into
two layers. The top layer consists of 9 sub-spaces, while the bottom
layer, excluding the obstacle itself, has 8 sub-spaces, resulting
in a total of 17 subspaces. UAV swarms, as typical
multi-agent systems, rely on cooperation among individual agents for
collision avoidance \cite{8439015} and the implementation of optimal
control \cite{9718582} to effectively address UAV conflict avoidance.
Another method to achieve collision-free flights between UAVs is to
define the flight range of the evading UAV as a sudden obstacle.
Sudden obstacles appear within the same sub-airspace of UAVs, such
as birds and non-cooperative UAVs. In Fig. \ref{fig:4}, different
from the static obstacles, the sudden obstacles are divided into three
layers, with each layer respectively containing 9, 8, and 9 subspaces,
and in total 26 subspaces. The method for calculating the distance
between the UAV and the sudden obstacle is the same as the static
obstacle. The size and position of the sudden obstacle may change,
and this model facilitates observation and dynamic adjustment of its
range with flexibility. 

In the trajectory planning process, obstacles are represented by their
closest position to the airspace origin, with a length in the x-axis,
y-axis, and z-axis. For example, ((2, 2, 0), 2, 3, 4) indicates that
the obstacle's closest position to the airspace origin is (2, 2, 0),
with a length of 2 meters in the x-axis, a length of 3 meters in the
y-axis, and a length of 4 meters in the z-axis. The difference between
sudden obstacles and static obstacles is that the coordinates of a
sudden obstacle can have a z-value greater than 0, and the rest of
the representation is the same as static obstacles.

\begin{figure}
\centering

\includegraphics[width=3.5in]{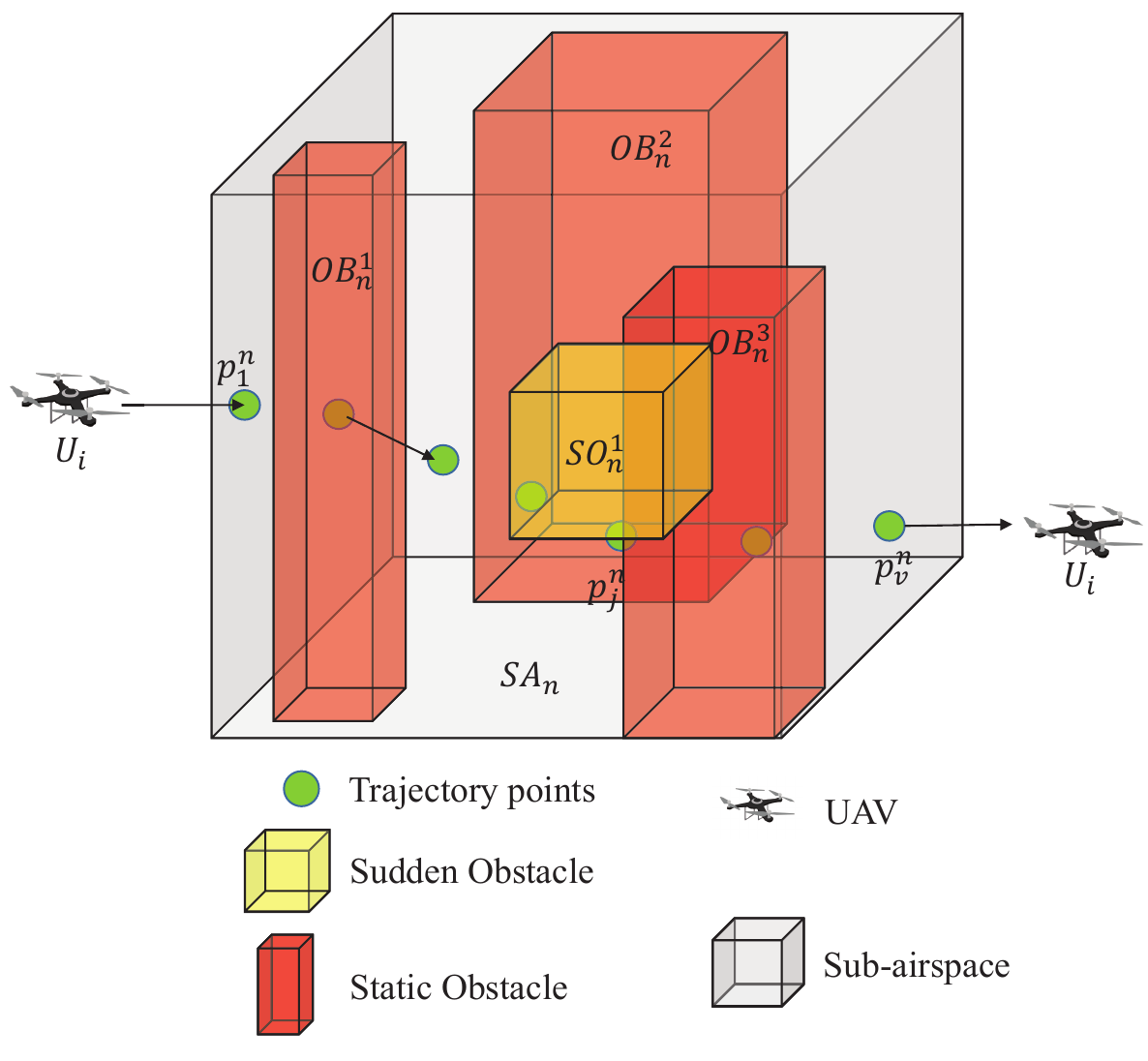}

\vspace{-3mm}

\caption{\label{fig:5}Trajectory planning in sub-airspace.}

\vspace{-5mm}
\end{figure}

The trajectory planning of UAV $U_{i}$ in sub-airspace $SA_{n}$
is shown in Fig. \ref{fig:5}. The red cuboid $\boldsymbol{OB}{OB_{n}^{1},\ OB_{n}^{2}\ \mathrm{and}\ OB_{n}^{3}}$
represents static obstacle, while the yellow cuboid represents a sudden
obstacle $SO_{n}^{1}$, which is already appeared before $U_{i}$
entering $SA_{n}$. The green points $\boldsymbol{P_{n}=}{p_{1}^{n},\ p_{2}^{n},\ ...,p_{j}^{n},\ ...,\ p_{J}^{n}}$
are the planned trajectory points for $U_{i}$ within the sub-airspace
$SA_{n}$. Trajectory $p_{j}^{n}$ point in $p_{n}$ consists of three-dimensional
coordinates $p_{j(x)}^{n}$, $p_{j(y)}^{n}$, and $p_{j(z)}^{n}$.

The vectors between two adjacent trajectory points in the $x$, $y$,
and $z$ directions are shown in formula (\ref{eq:6}) to formula
(\ref{eq:8}). The horizontal vector and the distance between two
adjacent trajectory points are represented by formula (\ref{eq:9})
and formula (\ref{eq:10}), respectively. The turning angle $TA_{n}^{i}$
and the pitch angle $PA_{n}^{i}$ of $U_{i}$ in $SA_{n}$ are obtained
through formula (\ref{eq:11}) and formula (\ref{eq:12}).

\begin{align}
\textrm{} & TA_{n}^{i}=arccos(\frac{\overrightarrow{p_{j+1(xy)}^{n}}\cdot\overrightarrow{p_{j(xy)}^{n}}}{\mid\overrightarrow{p_{j+1(xy)}^{n}}\mid\cdot\mid\overrightarrow{p_{j(xy)}^{n}}\mid}),\label{eq:6}\\
 & PA_{n}^{i}=arcsin(\frac{q_{j+1(z)}^{n}}{L_{j}^{j+1}}),\label{eq:7}\\
 & \overrightarrow{p_{j(xy)}^{n}}=(q_{j(x)}^{n},\;q_{j(y)}^{n}),\label{eq:8}\\
 & L_{j}^{j+1}=\sqrt{(q_{j+1(x)}^{n})^{2}+(q{}_{j+1(y)}^{n})^{2}+(q{}_{j+1(z)}^{n})^{2}},\label{eq:9}\\
 & q_{j(x)}^{n}=p_{j(x)}^{n}-p_{j-1(x)}^{n},\label{eq:10}\\
 & q_{j(y)}^{n}=p_{j(y)}^{n}-p_{j-1(y)}^{n},\label{eq:11}\\
 & q_{j(z)}^{n}=p_{j(z)}^{n}-p_{j-1(z)}^{n}.\label{eq:12}
\end{align}

$U_{i}$ evaluates the cost of the planned trajectory in the sub-airspace
through formula (\ref{eq:4}):

\begin{equation}
CSA_{n}^{i}=k3\cdotp(\frac{k5}{\sum_{j=1}^{O_{n}}lob_{n}^{j(e)}}+\frac{k6}{\sum_{j=1}^{M_{n}}lso_{n}^{j(e)}})+k4\cdotp\sum_{j=1}^{q-1}L_{j}^{j+1},\label{eq:4}
\end{equation}

where $CSA_{n}^{i}$ is the cost function of $U_{i}$'s trajectory
in $SA_{n}$. In formula (\ref{eq:4}), $O_{n}$ denotes the number
of static obstacles in $SA_{n}$, $M_{n}$ denotes the number of sudden
obstacles, $lob_{n}^{j(e)}$ indicates the distance between $j$th
trajectory point of $U_{i}$ and the $e$th static obstacle, and $lso_{n}^{j(e)}$
denotes the distance between $j$th trajectory point of $U_{i}$ and
the $e$th sudden obstacle. $J$ is the number of trajectory points
of $U_{i}$ in $SA_{n}$. $L_{j}^{j+1}$ denotes the distance between
adjacent trajectory points $p_{j}^{n}$ and $p_{j}^{n+1}$. $k3$,
$k4$, $k5$, and $k6$ are coefficients, and if the number of static
obstacles or sudden obstacles is 0, the parameters $k5$ or $k6$
are set as 0, respectively.

The positions of the trajectory points in $P_{n}$ are optimized in
problem $\mathscr{P}1$ to find a trajectory with the minimum cost:

\begin{align}
\mathscr{P}1:\;\underset{\boldsymbol{P_{n}}}{\textrm{min}}\  & CSA_{n}^{i}\label{eq:5}\\
\textrm{s.t.}\  & C1:L_{j}^{j+1}<l_{max},\nonumber \\
 & C2:\sum_{j=1}^{q-1}L_{j}^{j+1}<L_{max},\nonumber \\
 & C3:-TA_{max}^{i}<TA_{n}^{i}<TA_{max}^{i},\nonumber \\
 & C4:-PA_{max}^{i}<PA_{n}^{i}<PA_{max,}^{i}\nonumber \\
 & C5:SA_{n(min)}^{x}<p_{j(x)}^{n}<SA_{n(max)}^{x},\nonumber \\
 & C6:SA_{n(min)}^{y}<p_{j(y)}^{n}<SA_{n(max)}^{y},\nonumber \\
 & C7:SA_{n(min)}^{z}<p_{j(z)}^{n}<SA_{n(max)}^{z}.\nonumber 
\end{align}
In constraint $C1$, $l_{max}$ denotes the maximum length between
adjacent trajectory points, and in constraint $C2$, $L_{max}$ denotes
the maximum length of the trajectory in $SA_{n}$. In constraints
$C3$ and $C4$, $TA_{max}$ and $PA_{max}$ denote the maximum turning
angle and pitch angle between adjacent trajectory points, respectively,
while $TA_{min}$ and $PA_{min}$ denote the minimum turning angle
and pitch angle between adjacent trajectory points. Constraints $C5$,
$C6$, and $C7$ denote the position range constraints that each trajectory
point of the UAV needs to keep in safe ranges.

\begin{figure}
\centering

\includegraphics[width=3.5in]{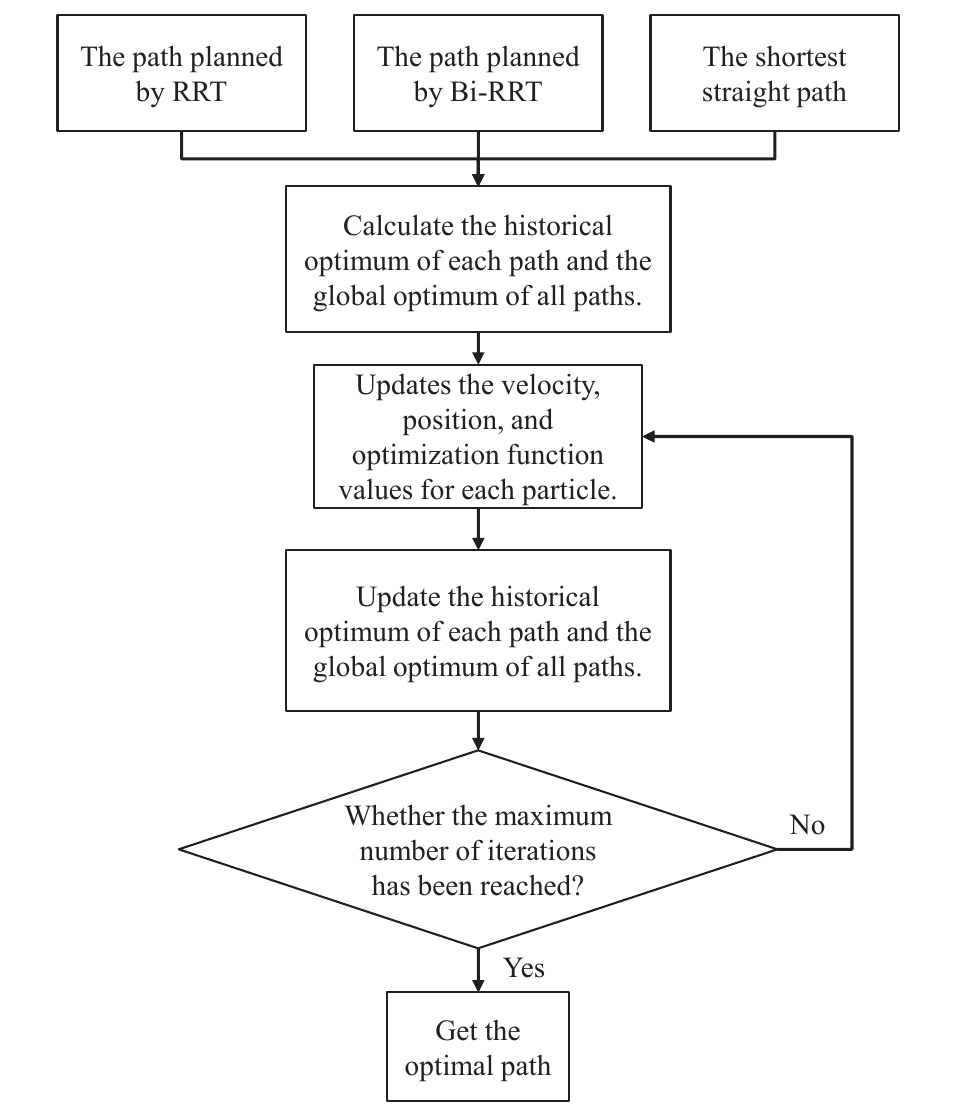}

\vspace{-3mm}

\caption{\label{fig:6}The process of PSO-RRT.}

\vspace{-5mm}
\end{figure}

In the process of trajectory planning, it is necessary to consider
both the distance between UAVs and obstacles and the length of the
trajectory. The length of the trajectory indicates the energy consumption
during UAV flight, while the distance between UAVs and obstacles represents
the flight safety. PSO is an optimization algorithm inspired by the
foraging behavior of birds. It has the fast convergence speed and
simple implementation, which is effective in solving optimization
problems. RRT is leveraged to find trajectories with no conflict within
a specified range. It has fast computation speed and generates random
trajectories. Bi-RRT is a variant of the RRT algorithm which adds
a greedy mechanism in searching progress, enabling faster calculation
of collision-free trajectories with shorter trajectory lengths. However,
Bi-RRT makes the trajectory closer to the obstacles. In order to make
full use of the advantages of PSO and Bi-RRT, we formulate PSO-RRT
to achieve quickly locating the waypoints of UAVs while keeping them
away from obstacles. Fig. \ref{fig:6} illustrates the proposed PSO-RRT
algorithm for sub-airspace trajectory planning based on PSO, RRT,
and Bi-RRT. Multiple RRT and Bi-RRT trajectories are planned within
sub-airspace $SA_{n}$, which serve as input data for PSO. An additional
trajectory connecting the starting point to the endpoint is added
as another set of input data to prevent the RRT algorithm from missing
the optimal obstacle-free trajectory. The input data is optimized
employing the PSO, updating the individual best positions and global
best positions of trajectory points in each UAV trajectory during
each iteration. The iteration process continues until the entire optimization
function converges or the maximum number of iterations is reached.
PSO-RRT can make full use of the randomness of RRT, providing greater
diversity in the input data. Bi-RRT provides trajectory data with
shorter lengths for the optimization process. PSO quickly optimizes
the input data, adjusts the position of trajectory points, and finds
the UAV trajectory with the minimum cost within $SA_{n}$. 

There are two situations in PSO-RRT when dealing with sudden obstacles.
Firstly, when PSO-RRT is conducting trajectory planning, a sudden
obstacle has already appeared. In such case, the treatment of sudden
obstacles is the same as static obstacles. Secondly, while the UAV
is flying within the sub-airspace, a sudden obstacle appears against
the planned trajectory. In this situation, it is necessary to re-plan
the trajectory within the sub-airspace to avoid the conflict.

\begin{figure}
\centering

\includegraphics[width=3.5in]{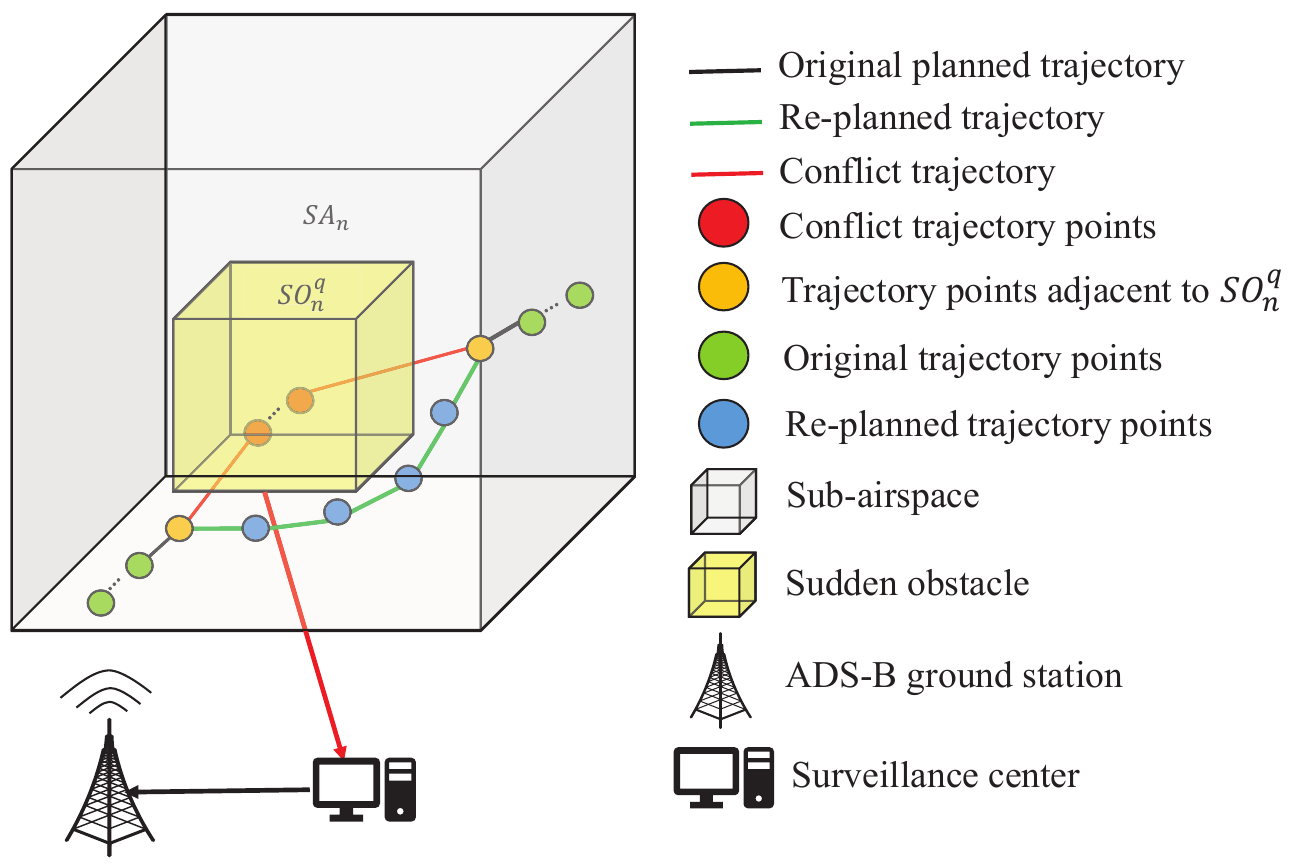}

\vspace{-3mm}

\caption{\label{fig:7}Trajectory re-planning for sudden obstacles in $SA_{n}$.}

\vspace{-5mm}\vspace{-1mm}
\end{figure}

As shown in Fig. \ref{fig:7}, when UAV $U_{i}$ is flying within
$SA_{n}$ according to the pre-planned trajectory points, the surveillance
center detects the occurrence of a sudden obstacle $SO_{n}^{q}$ within
$SA_{n}$. The surveillance center transmits the information of $SO_{n}^{q}$
to the ground station and broadcasts it to airspace via ADS-B. Upon
receiving this message, $U_{i}$ in $SA_{n}$ calculates whether there
is a conflict with the planned trajectory of $SO_{n}^{q}$. If there
is no conflict, $U_{i}$ continues to fly according to the current
trajectory. If there is a conflict with $SO_{n}^{q}$, a trajectory
re-planning is required. In Fig. \ref{fig:7}, the red trajectory
points represent the points that $U_{i}$ cannot reach due to conflicts,
the yellow trajectory points are the two original trajectory points
closest to the sudden obstacle $SO_{n}^{q}$, the green trajectory
points are the original trajectory points planned before entering
$SA_{n}$, and the blue trajectory points are the new trajectory points
generated by re-planning. $U_{i}$ leverages the two yellow trajectory
points as the starting and ending points for trajectory re-planning
in $SA_{n}$, and utilizes Bi-RRT to generate new blue trajectory
points to avoid conflicts with $SO_{n}^{q}$.

\section{Simulation Results\label{sec:4}}

\subsection{Trajectory planning in sub-airspace}

The parameter settings in $SA_{n}$ are shown in Table \ref{tab:1},
and the parameter settings of the PSO-RRT algorithm are shown in Table
\ref{tab:2}.

\begin{table}
\begin{centering}
\caption{\label{tab:1}Parameters of sub-airspace.}
\centering%
\begin{tabular}{|c|c|}
\hline 
Parameter & Value\tabularnewline
\hline 
\hline 
$SA_{n(min)}^{x}$, $SA_{n(max)}^{x}$ & 0, 200\tabularnewline
\hline 
$SA_{n(min)}^{y}$, $SA_{n(max)}^{y}$ & 0, 200\tabularnewline
\hline 
$SA_{n(min)}^{z}$, $SA_{n(max)}^{z}$ & 0, 50\tabularnewline
\hline 
$OB_{n}^{1}$ & ((40, 50 ,0), 50, 50, 100)\tabularnewline
\hline 
$OB_{n}^{2}$ &  ((20, 120, 0), 30, 30, 100)\tabularnewline
\hline 
$OB_{n}^{3}$ & ((150, 125, 0), 30, 30, 100)\tabularnewline
\hline 
$J$ & 10\tabularnewline
\hline 
$L_{max}$ & 400$\mathrm{m}$\tabularnewline
\hline 
$l_{max}$ & 40$\mathrm{m}$\tabularnewline
\hline 
$TA_{min}$, $TA_{max}$ & -60\textdegree , 60\textdegree{}\tabularnewline
\hline 
$PA_{min}$, $PA_{max}$ & -45\textdegree , 45\textdegree{}\tabularnewline
\hline 
\end{tabular}
\par\end{centering}
\vspace{1mm}

\vspace{-5mm}
\end{table}

\begin{table}
\caption{\label{tab:2}Parameters of PSO-RRT.}
\centering%
\begin{tabular}{|c|c|}
\hline 
Parameter & Value\tabularnewline
\hline 
\hline 
Trajectory planned by RRT & 15\tabularnewline
\hline 
Step size of RRT & 10m\tabularnewline
\hline 
Trajectory planned by Bi-RRT & 15\tabularnewline
\hline 
Step size of Bi-RRT & 10m\tabularnewline
\hline 
Smooth points & 5\tabularnewline
\hline 
Shortest straight trajectory & 1\tabularnewline
\hline 
Maximum number of iterations & 100\tabularnewline
\hline 
The weights of PSO & 0.8\tabularnewline
\hline 
The coefficients $c1$ and $c2$ of PSO & 1.4\tabularnewline
\hline 
The maximum velocity of particles & 2.5\tabularnewline
\hline 
$k3$ & 0.8\tabularnewline
\hline 
$k4$ & 0.2\tabularnewline
\hline 
$k5$ & 100\tabularnewline
\hline 
$k6$ & 100\tabularnewline
\hline 
\end{tabular}

\vspace{1mm}

\vspace{-5mm}
\end{table}

\begin{figure}[t]
\centering

\includegraphics[width=3.5in]{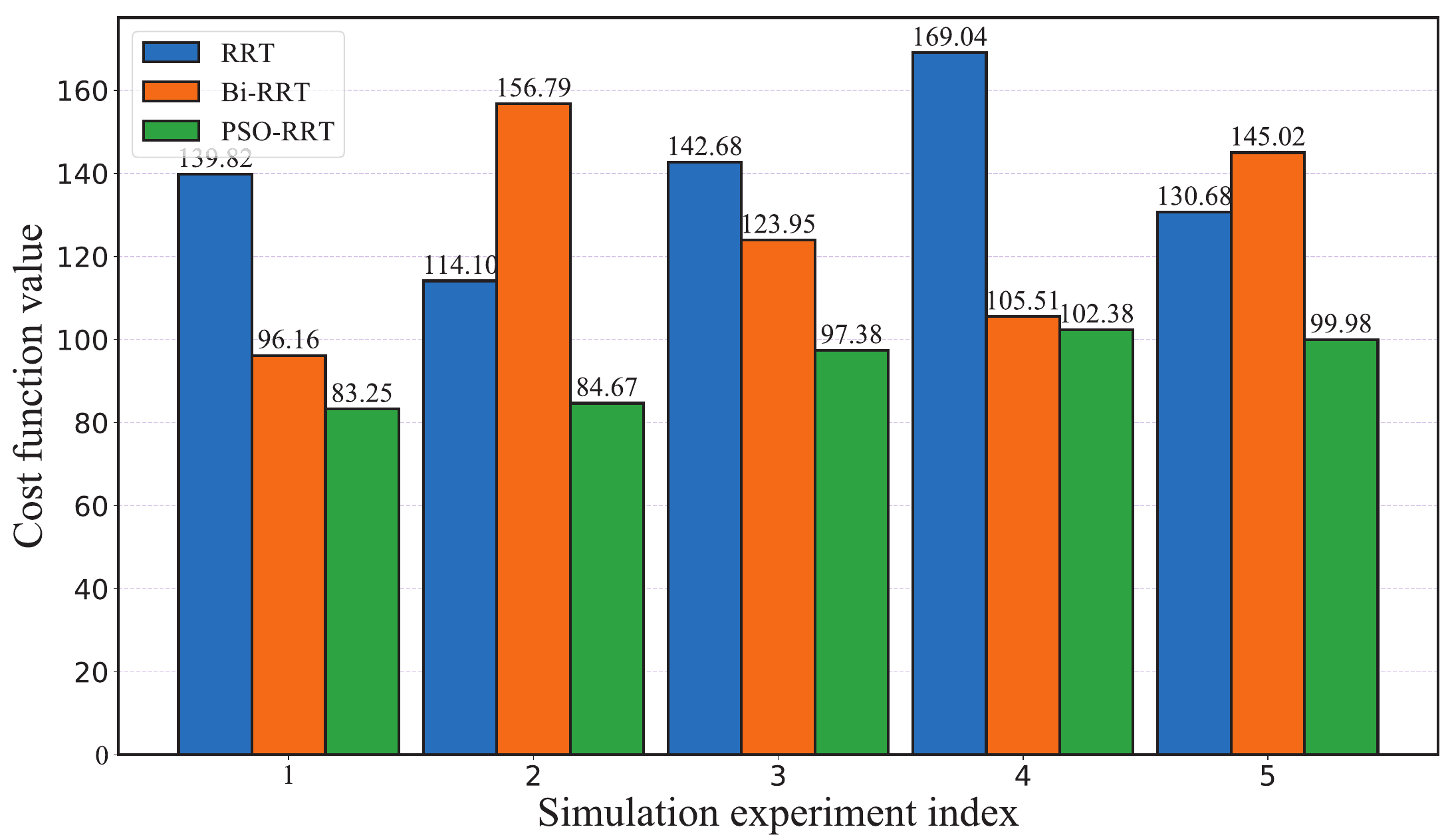}

\vspace{-3mm}

\caption{\label{fig:8}Cost function values of simulation experiments.}

\vspace{-5mm}

\end{figure}

The simulation results of cost function are shown in Fig. \ref{fig:8},
in detail, there is at least one obstacle between the starting point
and the endpoint of each esimulation. Each simulation leverages three
algorithms to calculate their cost function values. The trajectories
planned by RRT and Bi-RRT are the input data for PSO-RRT. It can be
seen that the cost function value of the trajectory planned in each
group of PSO-RRT is smaller than the cost function value of the trajectory
planned by the RRT and Bi-RRT algorithms. The reason lies in that
the trajectory planned by the two algorithms is leveraged as the input
data and the PSO is leveraged to adjust the optimization result. 

\begin{figure}
\centering

\subfloat[\label{fig:aaa}Top view of trajectory planning in sub airspace.]{\includegraphics[width=2.8in]{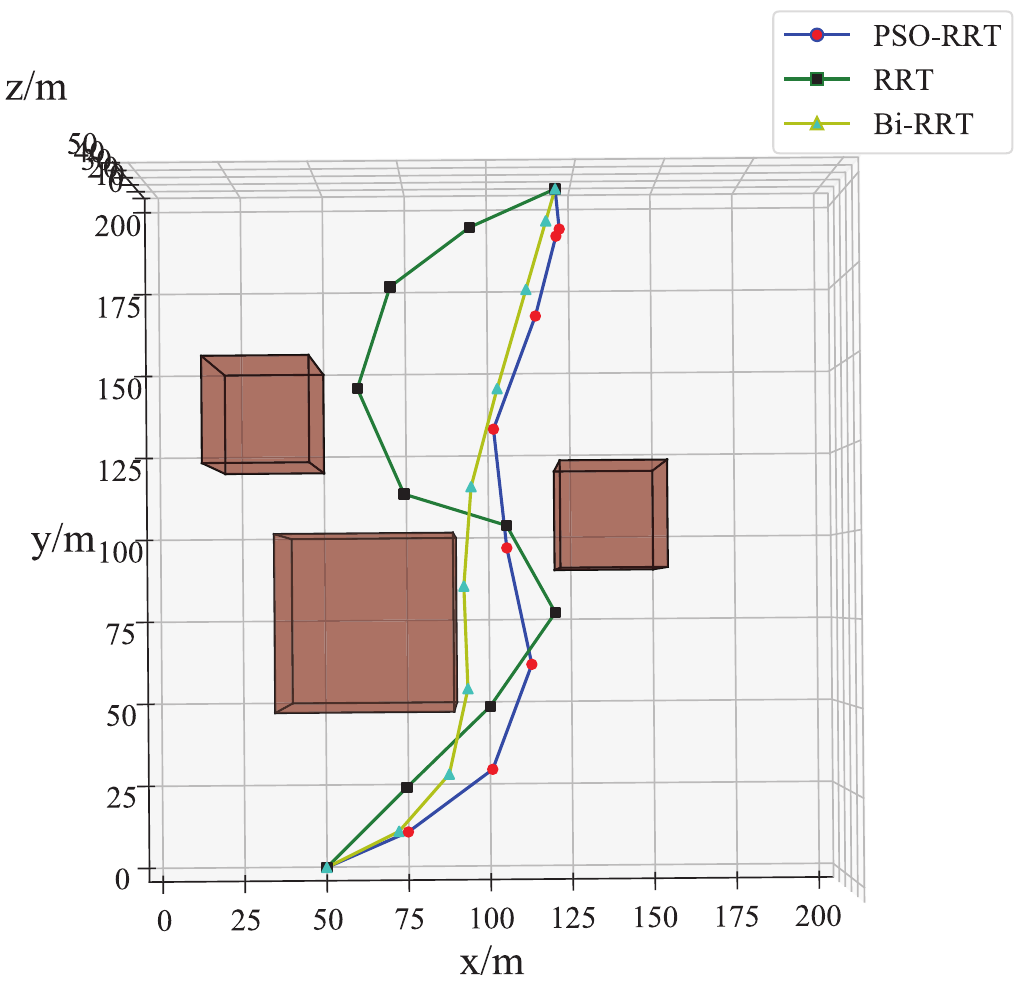}

\vspace{-3mm}

}

\vspace{-3mm}

\subfloat[\label{fig:bbb}Side view of trajectory planning in sub airspace.]{\includegraphics[width=2.8in]{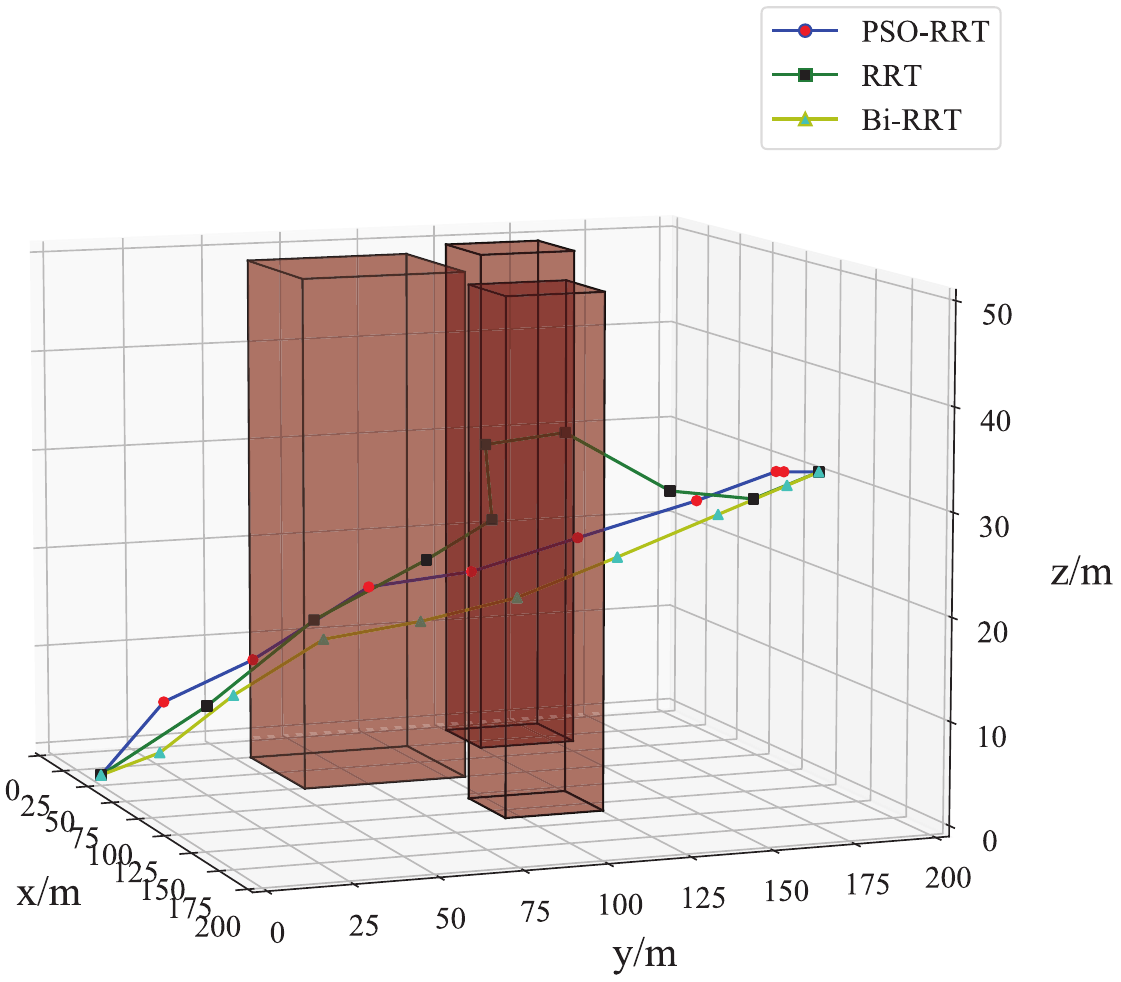}

\vspace{-3mm}

}

\caption{\label{fig:9}Trajectory planning in sub-airspace of simulation experiment
5. }

\vspace{-5mm}
\end{figure}

The top view and side view of the trajectory planned by the three
algorithms in $SA_{n}$ are respectively shown in Fig. \ref{fig:aaa}
and Fig. \ref{fig:bbb}. The green trajectory in Fig. \ref{fig:aaa}
is the smoothed trajectory planned by RRT. Since the algorithm planned
by RRT has strong randomness, it may lead to a longer trajectory length.
The yellow trajectory is planned by Bi-RRT. It is observed that due
to its own unique trajectory length greedy mechanism, the trajectory
is shorter than the other two algorithms. However, the shorter trajectory
brings a more radical planning strategy, which means the trajectory
is close to the obstacle, making $U_{i}$ easy to
collide with the obstacle if there exists an error
in positioning. The blue trajectory is the trajectory planned by PSO-RRT
for $SA_{n}$. Compared with the yellow and green trajectory, the
blue one maintains a shorter trajectory length and avoids two obstacles
between the second and sixth trajectory points, achieving a farther
distance from the obstacle, which means that PSO-RRT considers the
factors of security. Even if there is a deviation in positioning,
it can still safely reach the endpoint through static obstacles.

\begin{figure}
\centering

\includegraphics[width=3.5in]{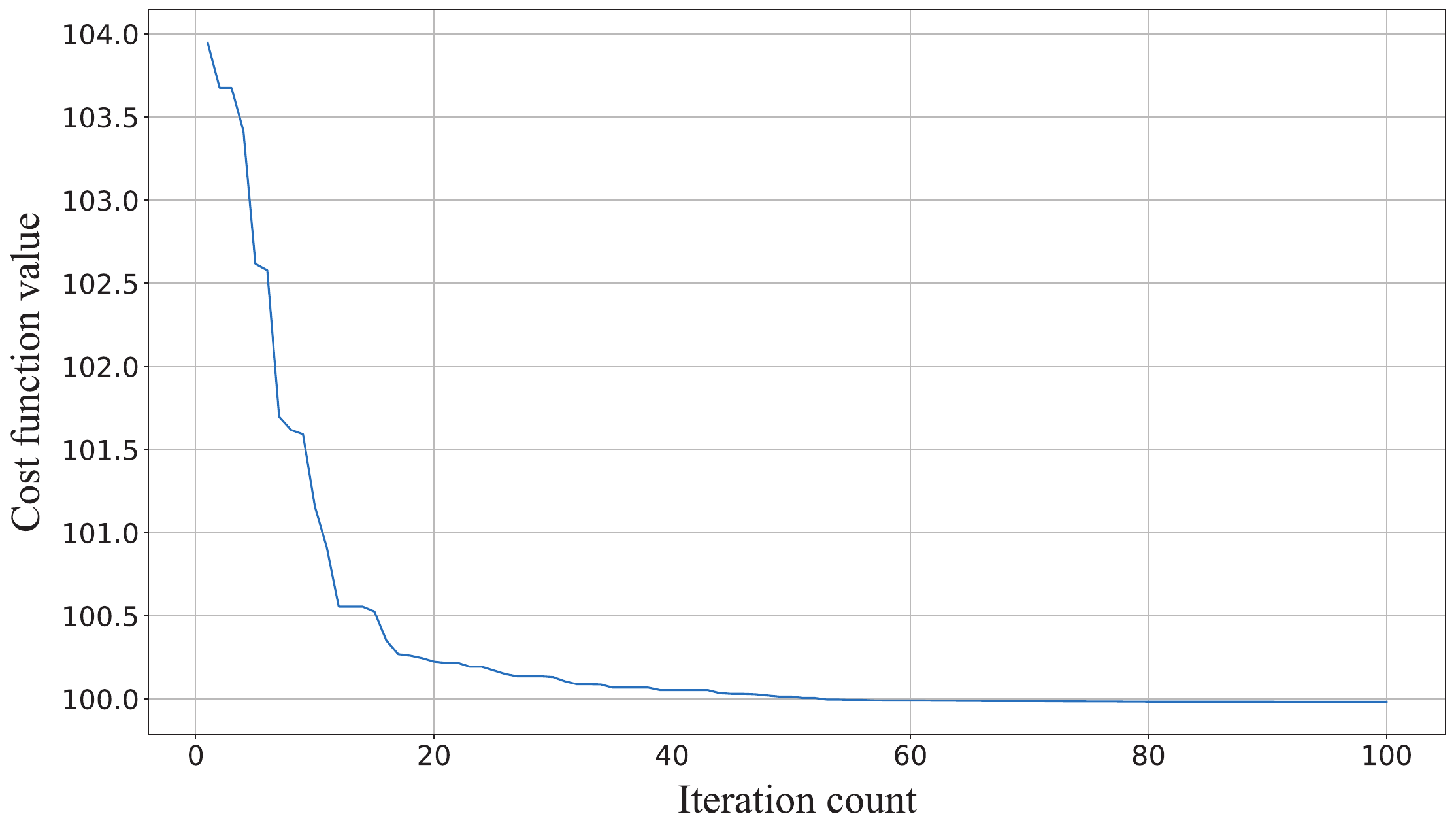}

\vspace{-3mm}

\caption{\label{fig:ccc}Cost function value of PSO-RRT.}

\vspace{-5mm}

\end{figure}

Fig. \ref{fig:ccc} shows the relationship between cost function value
and iterations in the planning process. At the beginning of the iteration,
the minimum cost function value of all trajectories is 103.94. With
the continuous increment of iterations, PSO constantly adjusts all
trajectories until the final cost function value converges to 99.98,
and the blue trajectory in Fig. \ref{fig:ccc} is obtained. Via the
change process of fitness value, it is concluded that the trajectory
obtained by final optimization must have a lower cost than the trajectory
before optimization.

\begin{figure}
\centering

\includegraphics[width=2.8in]{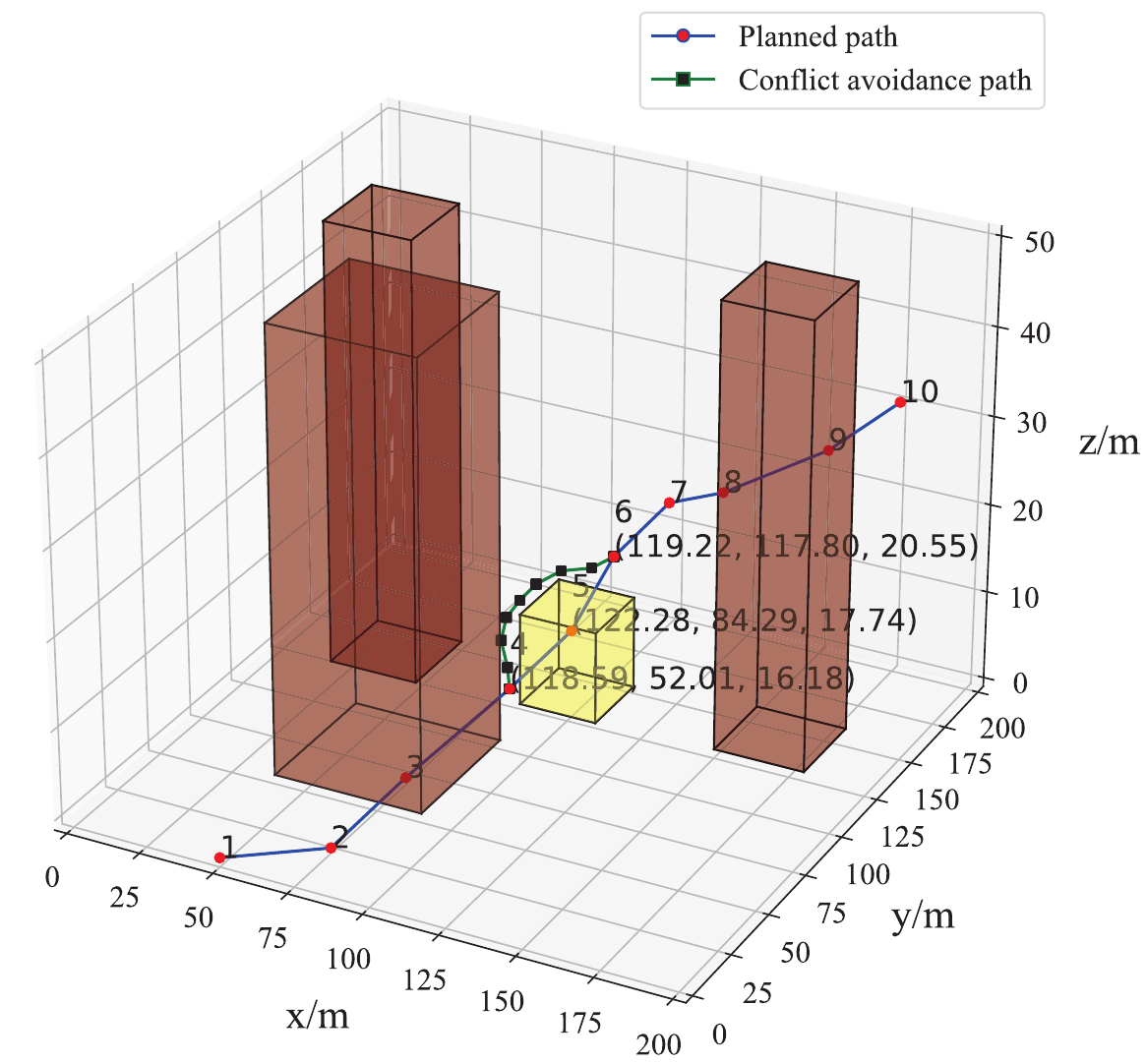}

\vspace{-3mm}

\caption{\label{fig:10}Trajectory re-planning in sub-airspace.}

\vspace{-5mm}
\end{figure}

In Fig. \ref{fig:10}, the UAV has planned the trajectory to be taken
for the current sub-airspace, but the monitoring center observes that
there exist sudden obstacles in the sub-airspace, and broadcasts the
obstacle information to the airspace through the ADS-B ground station.
After receiving the ADS-B information, the UAV in the sub-airspace
detects that the fifth trajectory point is located in the sudden obstacle,
resulting in the failure of the trajectory between the fourth trajectory
point and the sixth trajectory point. Therefore, the sudden obstacle
is added to the static obstacle list. Taking the coordinates of the
fourth trajectory point as the starting point and the coordinates
of the sixth trajectory point as the endpoint, Bi-RRT is leveraged
to quickly re-plan the trajectory, and the re-planned trajectory is
smoothed to replace the original conflict trajectory. When the UAV
is located at the fourth trajectory point, the new trajectory is executed
to avoid the yellow sudden obstacle.

\subsection{Trajectory planning among sub-airspaces}

After the whole airspace is divided, the trajectory planning is carried
out to verify the effect of sliding window and attraction mechanism.
The parameter settings of the experiment are shown in Table \ref{tab:3}.

\begin{table}[t]
\caption{\label{tab:3}Parameters of trajectory planning in airspace $A$.}
\centering

\begin{tabular}{|c|c|}
\hline 
parameter & value\tabularnewline
\hline 
\hline 
The lengths of $A$ in Three dimensions & 100,0m, 100,0m, 250m\tabularnewline
\hline 
$A_{x}$, $A_{y}$, $A_{z}$ & 5, 5, 5\tabularnewline
\hline 
Starting point & (0, 0, 0)\tabularnewline
\hline 
endpoint & (750, 900, 80)\tabularnewline
\hline 
Speed of UAVs & 5m/s\tabularnewline
\hline 
The number of obstacles & 75\tabularnewline
\hline 
Obstacle height range & {[}25, 240{]}\tabularnewline
\hline 
$k1$ & 0.01\tabularnewline
\hline 
$k2$ & 0.99\tabularnewline
\hline 
\end{tabular}

\vspace{1mm}

\vspace{-5mm}

\end{table}

\begin{figure}
\centering

\subfloat[\label{fig:111}Top view of trajectory planning in airspace $A$.]{\includegraphics[width=2.8in]{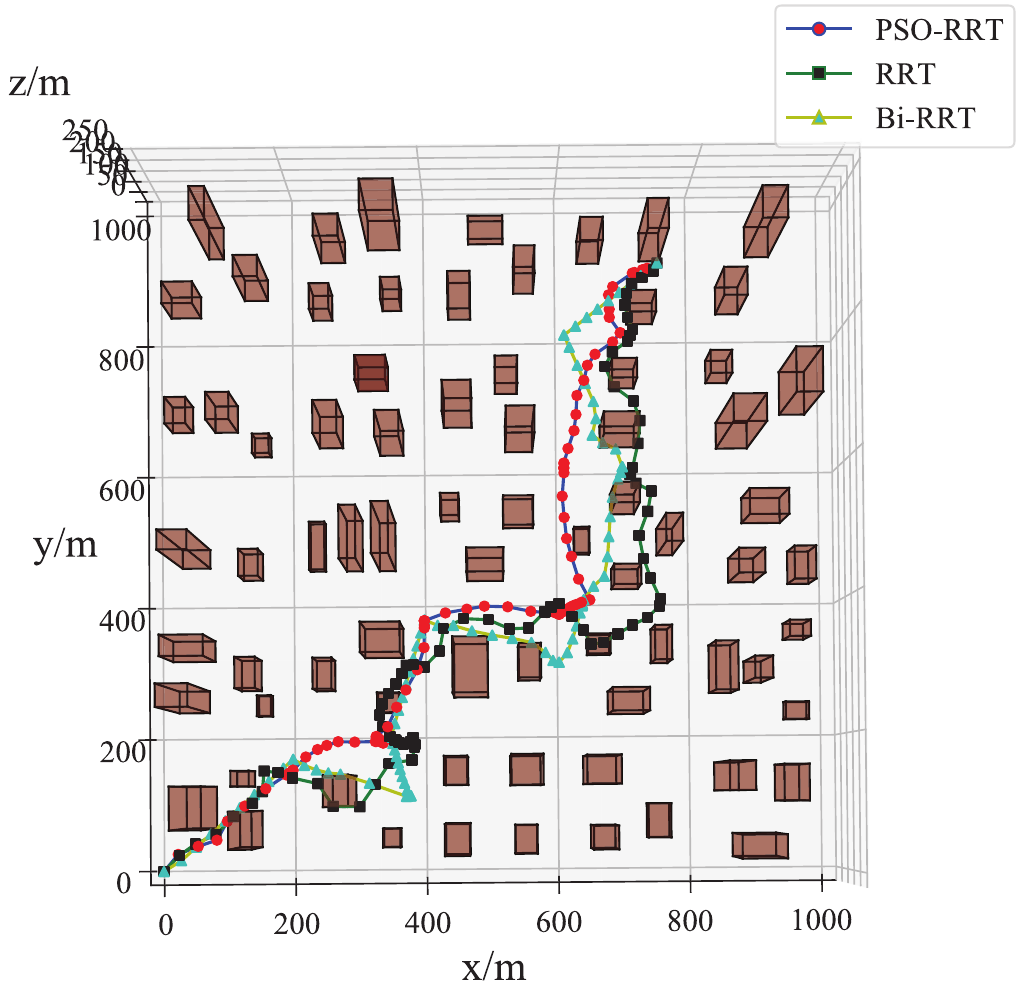}
\vspace{-3mm}

}

\vspace{-3mm}

\subfloat[\label{fig:222}Side view of trajectory planning in airspace $A$.]{\includegraphics[width=2.8in]{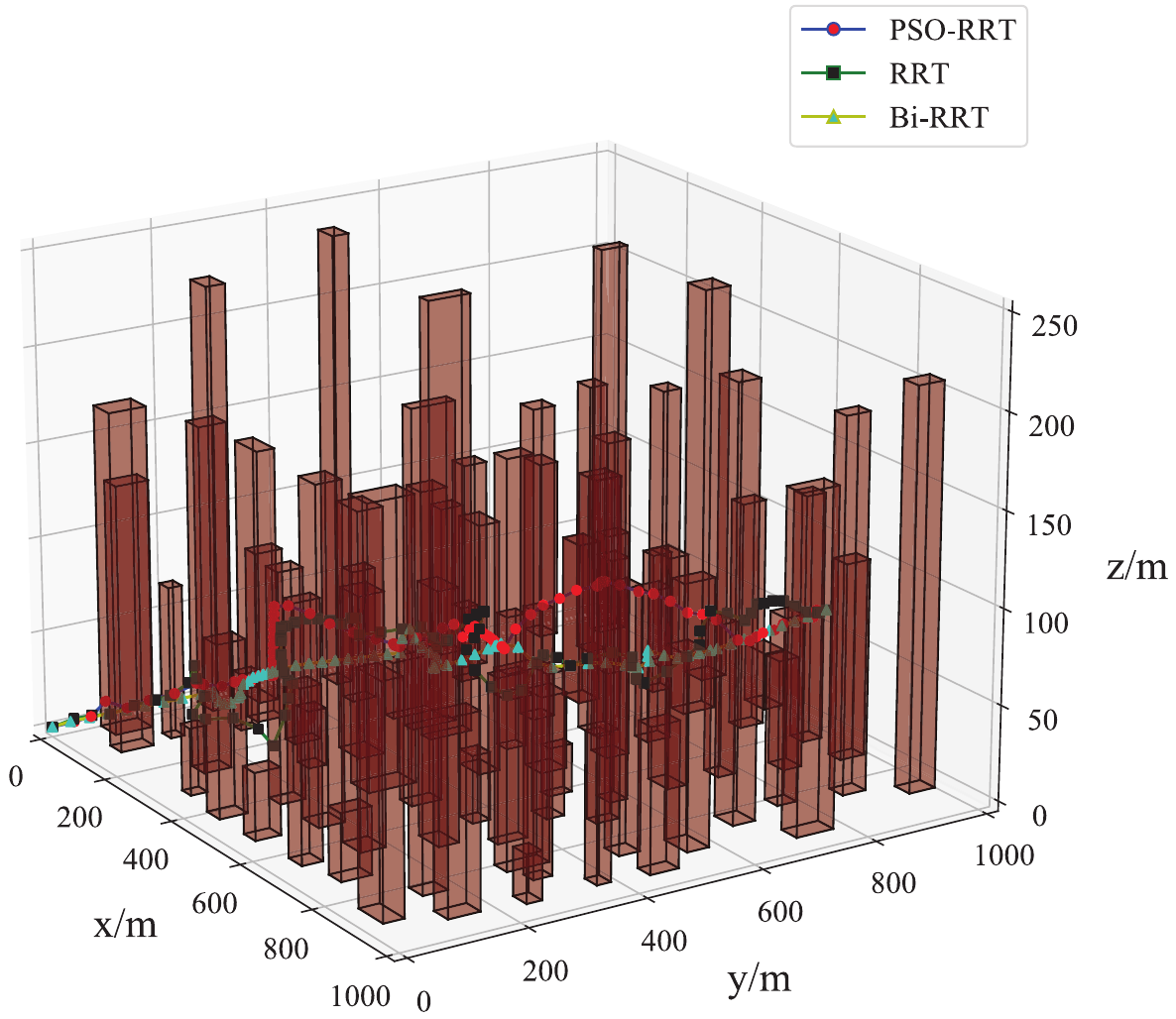}

\vspace{-3mm}

}

\caption{\label{fig:11}Trajectory planning in airspace $A$ with different
algorithm.}

\vspace{-5mm}
\end{figure}

The obstacle distribution of the whole airspace $A$ in the experimental
setting is shown in Fig. \ref{fig:11}. SSP and three trajectory planning
algorithms are employed with same environment settings. The sub-airspace
in $CS$ is: ($SA_{1}$, $SA_{2}$, $SA_{27}$, $SA_{32}$, $SA_{33}$,
$SA_{34}$, $SA_{39}$, $SA_{44}$, $SA_{49}$). The black trajectory
point is the trajectory planned by SSP and RRT, and the blue trajectory
point is the trajectory planned by SSP and Bi-RRT. It is observed
that the distance between the trajectory point and the obstacle is
very close in the whole trajectory. The red point is the trajectory
planned by SSP and PSO-RRT, and the trajectory point maintains a large
distance from the obstacle.

\begin{figure}
\centering

\includegraphics[width=3.5in]{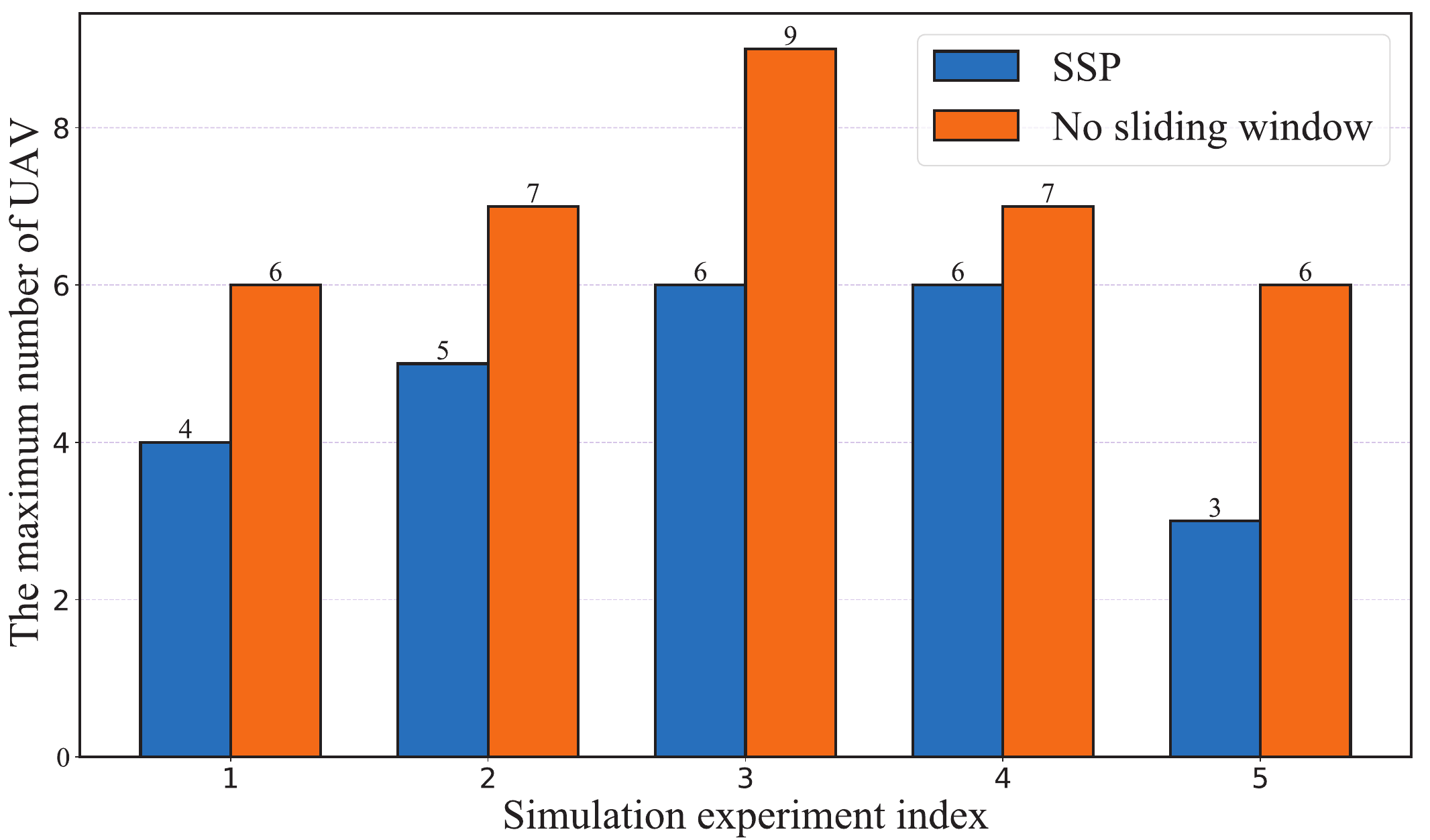}

\vspace{-3mm}

\caption{\label{fig:12}The maximum number of UAVs in simulation experiments.}

\vspace{-5mm}

\end{figure}

In Fig. \ref{fig:12}, we set the number of UAVs in airspace $A$
as $50$, and randomly generate $5$ sets of starting points and endpoints
for these UAVs. The trajectory between the starting point and the
endpoint includes at least $5$ sub-airspaces. Each set of simulations
only leverages SSP and no sliding window method for UAVs. The trajectory
of no sliding window is determined before take off. In the simulation,
the maximum number of UAVs in the sub-airspace of the two trajectory
planning methods during the entire UAV flight is recorded. It is observed
that the maximum number of UAVs in the sub-space domain of SSP is
smaller than that of trajectory planning without sliding window, which
further guarantees the safety.

\begin{figure}
\centering

\includegraphics[width=3.5in]{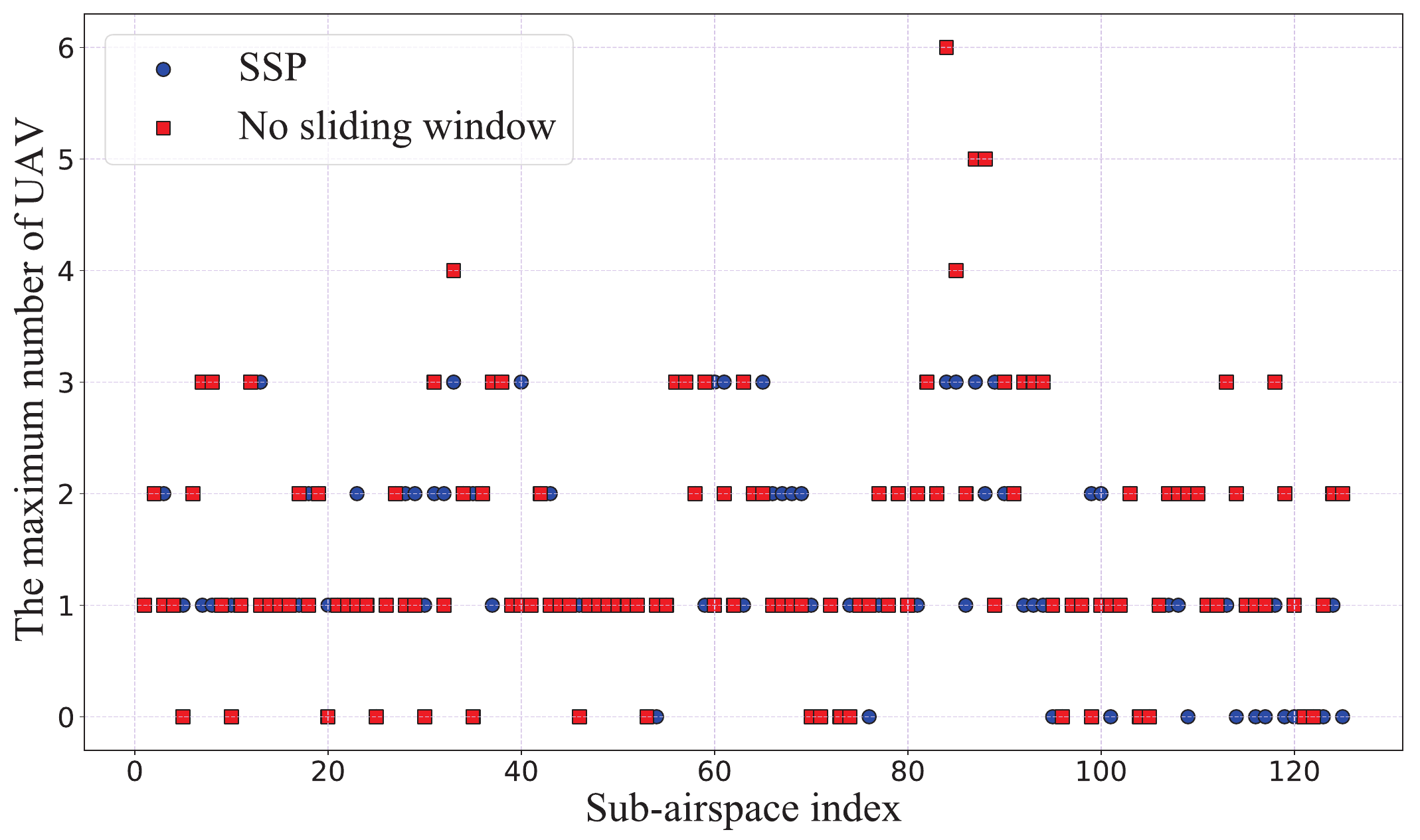}

\vspace{-3mm}

\caption{\label{fig:13}The number of UAVs in each sub-airspace in simulation
index $5$.}

\vspace{-5mm}
\end{figure}

Fig. \ref{fig:13} is the distribution of the maximum number of UAVs
in $SA_{1}$ to $SA_{125}$ of the fifth index of simulations in Fig.
\ref{fig:12}, leveraging SSP and trajectory planning without sliding
window. The maximum number of UAVs in the sub-airspace of SSP is 3,
while the maximum number of UAVs in the sub-airspace without sliding
window is 6, which are larger than the results of SSP. The reason
is that the sliding window adjusts the subsequent sub-airspace trajectory
according to the number of UAVs in the whole sub-airspace broadcast
by ADS-B ground station when the UAV enters the new sub-airspace,
which effectively reduces the maximum number of UAVs in each sub-airspace.
The trajectory of no sliding window is fixed before the UAV takes
off, so when the number of UAVs in the airspace is very large. The
subsequent trajectory cannot be adjusted.

\begin{figure}
\centering

\includegraphics[width=3.5in]{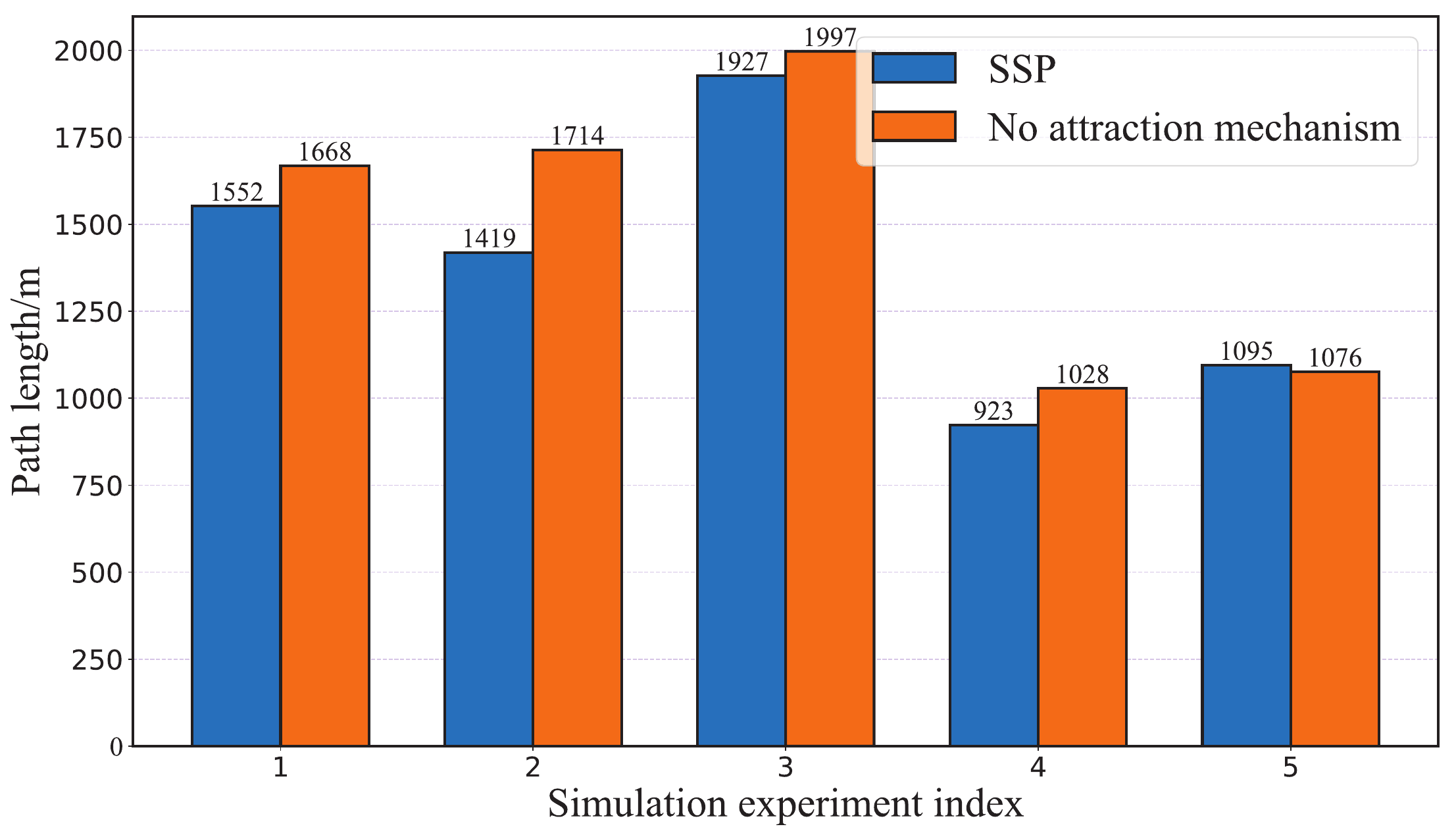}

\vspace{-3mm}

\caption{\label{fig:14}Trajectory lengths of different trajectory planning
methods in airspace $A$.}

\vspace{-5mm}
\end{figure}

In Fig. \ref{fig:14}, five sets of simulations are carried out and
the first four are randomly generated. There are at least five sub-airspaces
between the starting point and the endpoint. The fifth simulation
\textcolor{blue}{is} set specially, and its starting point and the
endpoint are respectively $SA_{1}$ and $SA_{5}$ with no direction
change. As shown in Fig. \ref{fig:14}, in the first four groups of
simulations, due to the change of direction among the sub-airspace
trajectories, the attraction mechanism reduces the range of the sub-airspace
endpoint in the trajectory planning process and reduces the total
trajectory length. However, the trajectory planning without the attraction
mechanism has a longer trajectory length because the endpoint in the
sub-airspace is completely randomly selected. In the fifth simulation,
since there is no change in the direction of $CS$, the attraction
mechanism fails, which is the same as the completely random search
for the endpoint of the sub-airspace. Therefore, the length of the
trajectory in the sub-airspace with the attraction mechanism is longer.

\section{Conclusion and Future Work\label{sec:5}}

In this paper, in order to enhance the information acquisition and
environmental perception capabilities, UAVs in low-altitude urban
areas are equipped with ADS-B devices to achieve high-frequency information
exchange. In order to enhance the safety and efficiency of UAVs, we
divide the low-altitude urban airspace into multiple sub-airspaces,
and leverage ADS-B to continuously monitor flight for each sub-airspace.
On the basis of airspace division, we propose SSP algorithm based
on dynamic programming, sliding window and attraction mechanism to
conduct coarse-grained trajectory planning among sub-airspaces, and
we propose the PSO-RRT algorithm for trajectory planning in sub-airspaces.
The results of multiple simulations prove that the maximum number
of UAVs in sub-airspaces and the total length of trajectory are both
reduced by SSP. As for the trajectory planning in sub-airspace, the
PSO-RRT algorithms reduce the cost of trajectory compared with the
trajectory planned by RRT and Bi-RRT, which means the trajectory planned
by PSO-RRT simultaneously considers both safety and efficiency. \textcolor{black}{In
conclusion, the collision-free trajectory planning for UAVs within
the airspace has been successfully implemented by SSP and PSO-RRT
with ADS-B information.}\textcolor{blue}{{} }

To further investigate the real-time performance,
UAV peak values in airspace, and average computation time of the proposed
algorithm, we just constructed a set of ADS-B OUT and ADS-B IN devices
leveraging Raspberry Pi, positioning modules, and software-defined
radio equipment to test the algorithm's performance. In
addition, in future work, we will incorporate cooperation between
UAVs into UAV conflict avoidance considerations, and consider the
impact of NACv and NACp on ADS-B message reception, conducting more
practical experiments for validation using the ADS-B devices we have
constructed.

\bibliographystyle{IEEEtran}
\bibliography{123}
\end{CJK}
\end{document}